\documentclass[final,5p,times,twocolumn]{elsarticle}
 \biboptions{comma,sort&compress}
\usepackage{graphicx}
\usepackage{amsmath}
\usepackage{here}

\usepackage[dvips]{epsfig}

\usepackage[table]{xcolor}
\usepackage{xcolor}
\usepackage{soul}
\usepackage{ulem}
\usepackage{hyperref}

\definecolor{bluette}{rgb}{.2,.4,0}
\definecolor{salmon}{rgb}{.9,0.68,0.5}
\definecolor{motive}{rgb}{0.2,1,.5}
\definecolor{list}{rgb}{0.3,.8,.1}
\definecolor{moe}{rgb}{1,.7,.5}
\definecolor{mote}{rgb}{.7,.5,.6}
\definecolor{pisello}{rgb}{.1,1,0}
\definecolor{orange}{rgb}{1,.7,0}
\definecolor{oliva}{rgb}{.1,.5,0.3}
\definecolor{greenda}{rgb}{0,.3,.2}
\definecolor{greenli}{rgb}{0.5,.8,.0}
\definecolor{blueda}{rgb}{0,.1,.6}
\definecolor{purple}{rgb}{.7,.1,.2}
\definecolor{marrone}{rgb}{1,0.7,0}
\definecolor{pinky}{rgb}{1,0.8,0.8}
\definecolor{rose}{rgb}{1,0.4,0}

\def\oliva{\color{oliva}}

\def\purple{\color{purple}}
\def\blue{\color{blueda}}

\def\beq{\begin{equation}}
\def\eeq{\end{equation}}
\def\bea{\begin{eqnarray}}
\def\eea{\end{eqnarray}}
\def\bq{\begin{quote}}
\def\eq{\end{quote}}
\def\bm{\boldmath}

\def\nnb{\nonumber}
\def\ga{\left(}
\def\dr{\right)}

\def\lrar{\longrightarrow}

\def\nnb{\nonumber}
\def\la{\langle}
\def\ra{\rangle}
\def\nin{\noindent}
\def\ba{\vspace*{-0.2cm}\begin{array}}
\def\ea{\end{array}\vspace*{-0.2cm}}

\def\b{$\bullet~$}
\def\d{$\diamond~$}
\def\als{\alpha_s}

\def\gg2{ \la\alpha_s G^2 \ra}
\def\gg3{g^3f_{abc}\la G^aG^bG^c \ra}
\def\ggg4{\la\als^2G^4\ra}

\journal{Elsevier}

\begin{document}

\begin{frontmatter}

\title{ New Dynamics (?) of $J/\psi J/\psi$ and $\Upsilon\Upsilon$ families from QCD Laplace sum rules at NLO\,$^{1}$}    
 \author[label1,label2]{Stephan Narison\fnref{fn1}}
\address[label1]{Laboratoire
Univers et Particules de Montpellier, CNRS-IN2P3, 
Case 070, Place Eug\`ene
Bataillon, 34095 - Montpellier, France}
\address[label2]{Institute of High-Energy Physics (iHEPMAD), Univ. Antananarivo, Madagascar}

\fntext[fn0]{This paper is a Tribute to Professor Eduardo de Rafael, for his pioneering works on QCD spectral sum rules, who passed away during the preparation of this paper.}
\fntext[fn1]{Corresponding author}
  \ead{snarison@yahoo.fr}

 \author[label2]{Andry Rabemananjara}
\ead {achrisrab@gmail.com}
   
    \author[label2]{Davidson Rabetiarivony}
   \ead{rd.bidds@gmail.com}

\begin{abstract}
\nin
Motivated by the recent CMS data on a family  of $2^{++}$ states from $J/\psi J/\psi$ mass spectrum, 
we determine the masses of the $2^{++}$  $J/\psi J/\psi$ molecule and $cc\bar c\bar c$ like-four-quark states,  
by using {\it (for  the first time)} the two lowest QCD (inverse) Laplace Sum Rules (LSR) moment-ratios within the standard optimization criteria with respect to the LSR variable $\tau$ and to the QCD continuum threshold $t_c$ implemented by the $R_{P/C}$ requirement that the lowest ground state (Pole) contribution to the LSR is larger than the QCD continuum one. Including factorized NLO perturbative and LO $\la G^4\ra$ gluon condensate  contributions in the OPE, we obtain the {\it conservative  estimate} of the lowest ground state molecule mass\,: $M^{2^{++}}_{J/\psi J/\psi}=6521(69)$ MeV and four-quark state\,: $M^{2^{++}}_{cc\bar c \bar c}=6347(99)$ MeV with the couplings\,: $f^{2^{++}}_{J/\psi J/\psi}=96(14 )$ keV and $f^{2^{++}}_{cc\bar c \bar c}= 151(29)$ keV normalized as $f_\pi=93$ MeV. The masses of the first radial excitations are found to be\,:
$M^{2^{++}{\rm rad}}_{J/\psi J/\psi}=7987(251 )$ MeV and $M^{2^{++}{\rm rad}}_{cc\bar c \bar c}=7739(221)$ MeV indicating a mass-splitting about $2.5 (M_{\psi(2S)}-M_{J/\psi})$.  The corressponding decay constants are $f^{2^{++}{\rm rad}}_{J/\psi J/\psi}=248(70)$ keV and \,: $f^{2^{++}{\rm rad}}_{cc\bar c \bar c}=453(96) $ keV which are (unexpectedly) larger than the lowest ground state ones. These features may indicate some new dynamics compared to ordinary $\bar cc$ states.
 We extend the analysis to the $0^{++}$ states and to the $\Upsilon\Upsilon$ family where the results are compiled in Tables \,\ref{tab:psi-psi-0} and \,\ref{tab:upsilon-2}. We critically review some other QCD spectral sum rules (QSSR) results.  We  suggest that the $X(6600)$ might be (mainly) a molecule ground state. Using a two-component mass mixing scheme,  the $X(6900)$ and $X(7100)$ might be interpreted as a mixture of the $cc\bar c\bar c$ lowest ground state with its 1st radial excitation via an angle $\theta\approx 5^0$.

\end{abstract}
\end{frontmatter}
\section{Introduction}
\vspace*{-0.15cm}
QCD spectral sum rules (QSSR)
\`a la SVZ\,\cite{SVZa,SVZb,ZAKA} have been applied 
since 47 years\,\footnote{For reviews, see e.g\,\cite{SNB1,SNB2,SNB3,SNB26,SNB4,SNB5,SNREV15,SNREV10,IOFFEb,RRY,DERAF,BERTa,YNDB,PASC,DOSCH}.}  to study successfully the hadron properties (masses, couplings and widths) and to extract some fundamental QCD parameters  ($\alpha_s$, quark masses, quark and gluon condensates,...) from the QCD Lagrangian. In previous series of papers\,\cite{HEP18,SU3,QCD16,MOLE16,X5568,MOLE12,MOLE20}\,\footnote{For reviews on the use of QCD spectral sum rules for molecules and four-quark states, see e.g.\,\cite{MOLEREV,ZHUREV}.}, we have used the inverse Laplace transform (LSR)\,\cite{BELLa,BELLb,BECCHI,SNR,SNREV}  of  QSSR to predict the couplings and masses of different heavy-light molecules and tetraquarks states by including next-to-next nonleading order (N2LO) factorized perturbative (PT) corrections where we have emphasized the importance of these corrections for giving a meaning of the input heavy quark mass which plays an important role in the analysis though these corrections are (relatively) small in some channels within the $\overline{MS}$-scheme. However, this  (a posteriori) feature can justify the uses of the  $\overline{MS}$ running masses at LO often done in the current literature. 
In this paper, we test the assignement as $J/\psi J/\psi$ molecule and / or its four-quark state analogue for the  three exotic states $X(6600, 6900\,{\rm and}\, 7100)$ with $J^{PC}=2^{++}$ decaying into $J/\psi J/\psi$ recently found by CMS\,\cite{CMS}\,\footnote{LHCb\,\cite{LHCb}, ATLAS\,\cite{ATLAS} and CMS\,\cite{CMS1} also found some other not fully explained bumps in the region 6.2 to 7.4 GeV.}.  In so doing, we  use moments of Inverse Laplace sum rules (LSR) within the standard optimization criteria with respect to the LSR variable $\tau$ and the QCD continuum threshold $t_c$. We shall fix the subtraction scale at the optimal values $\mu=4.5$ GeV for the charm and  $\mu=7.25$ GeV for the bottom as deduced from our analysis of different molecules and four-quark states containing the charm quark\,\cite{HEP18,SU3,QCD16,MOLE16,X5568,MOLE12,MOLE20}. We shall include the contribution of the  factorized NLO perturbative and the one of LO dimension-eight gluon condensates in the OPE. To the previous stability criteria, we shall also request that the results will only be considered in the region where the lowest ground state pole contribution exceeds the QCD continuum one.

We shall also improve our previous related results in the $0^{++}$ scalar channel\,\cite{MOLE20} using a similar approach and extend the analysis to the bottom quark channel.


\begin{figure*}[hbt]
\vspace*{-3cm}
\begin{center}
\includegraphics[width=14.cm]{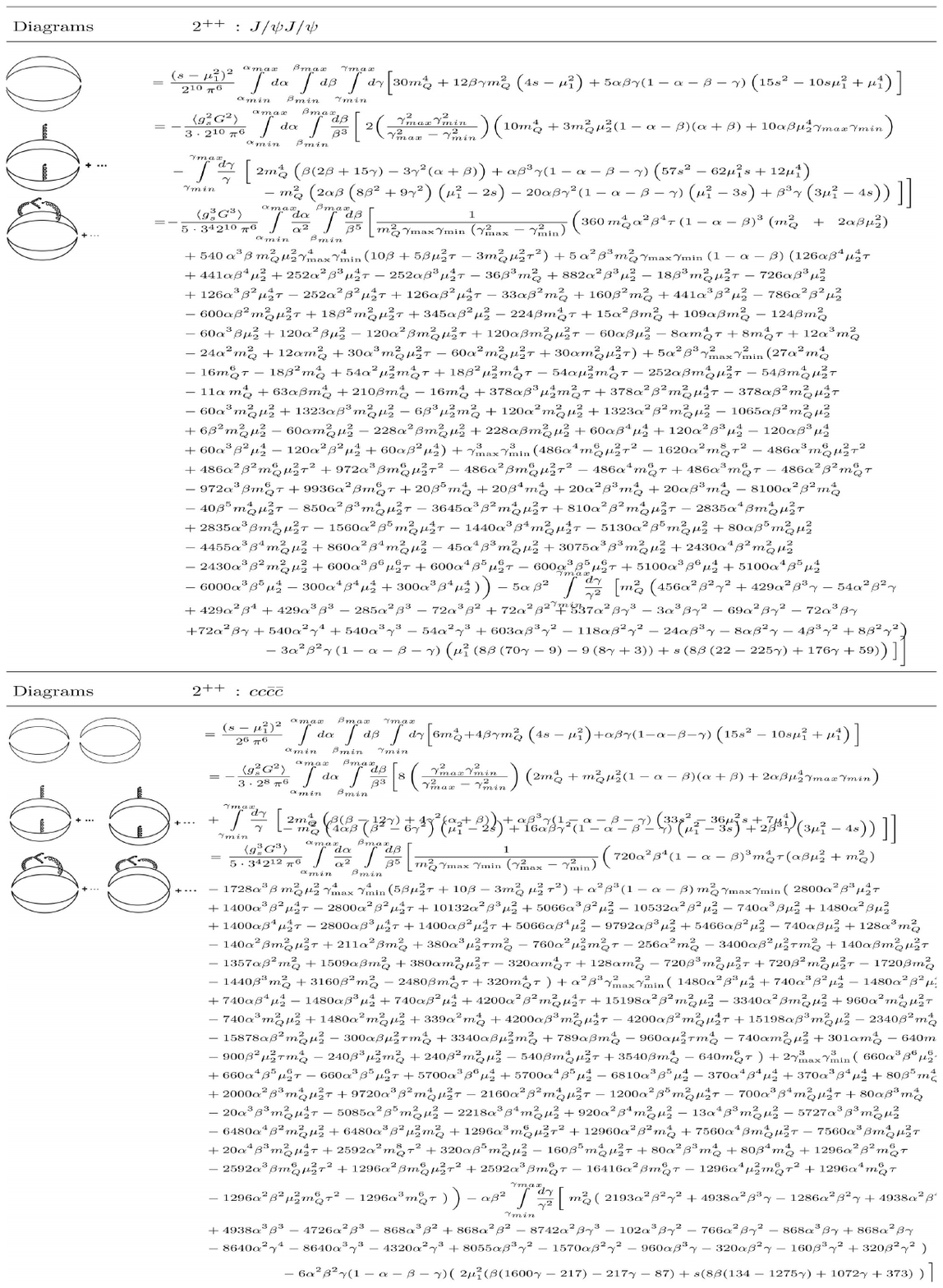}
\vspace*{-0.25cm}
\caption{\footnotesize  QCD expression of the $2^{++}$ $J/\psi J/\psi$ molecule two-point function to Lowest Order and up to dimension-six $\la G^3\ra $  condensate. }
\label{fig:qcd1}
\end{center}
\vspace*{-1cm}
\end{figure*} 
\section{The $2^{++}$ two-point correlator}
  We shall be concerned with the following QCD local interpolating current of dimension-six:
\bea
\hspace*{-0.3cm}{\cal O}^{\mu\nu}_{J/\psi J/\psi}(x)&\equiv& (\bar c\gamma^\mu c ) (\bar c\gamma^\nu c )(x)\, \nnb\\
{\cal O}^{\mu\nu}_{cc\bar c\bar c}(x)&\equiv&
\epsilon_{abc}\epsilon_{dec}
\left(
c_a^T C\gamma^\mu c_b
\right)
\left(
\bar c_d \gamma^\nu C \bar c_e^T
\right). 
\eea
\\
for the $J/\psi\, J\psi$ molecule and $cc\bar c \bar c$ four-quark states.  Among different choices of the four-quark dimension-six interpolating operators which should (in principle) mix each others under renormalization\,\cite{TARRACH2}, we choose the local diquark-antidiquark
current built from axial-vector diquark in the attractive color
antitriplet configuration  analogue to the $J/\psi J/\psi$ molecule current. 
 The corresponding two-point correlator is:
  \bea
 \hspace*{-0.4cm} \psi^{\mu\nu\rho\sigma}_X&=&i\hspace*{-0.1cm}\int\hspace*{-0.1cm} d^4x\, e^{iqx} \la 0|\vert {\cal T}{\cal O}^{\mu\nu}_{X}(x) {\cal O}^{\rho\sigma}_{X}(0)^\dagger\vert 0 \ra\nnb\\
  &=& \frac{1}{2} \ga \eta^{\mu\rho} \eta^{\nu\sigma}+ \eta^{\mu\sigma} \eta^{\nu\rho}-\frac{2}{3} \eta^{\mu\nu} \eta^{\rho\sigma}\dr \psi_X(q^2),\nnb\\
{\rm  with} \,\,\,\, \eta^{\mu\nu}& \equiv& g^{\mu\nu}-q^\mu q^\nu/q^2. 
  \eea
  The LO QCD expressions up to dimension-six condensates are given in Fig\,\ref{fig:qcd1}\,\footnote{We thank Raphael  Albuquerque for providing these results prior publication which we have checked up to the $\la G^2\ra$ contribution.} .  The limits of integration in $\alpha$, $\beta$ and $\gamma$ variables are\,:\\
  \vspace*{-0.3cm}
  { \footnotesize
\bea
  \hspace*{-2cm}\gamma_{_{\substack{max \\ min}}} \hspace*{-0.3cm} &=& \hspace*{-0.3cm} 
  \frac{1}{2} (1-\alpha -\beta) \left(~ 1 \:\pm\: \sqrt{1 + 
  \frac{4 m_Q^2 \alpha \beta}{(1 - \alpha - \beta) 
  \Big( m_Q^2(\alpha +\beta ) - \alpha  \beta  s \Big)}} ~\right) \nonumber\\
  \nonumber\\
  \hspace*{-2cm} \beta{_{\substack{max \\ min}}} \hspace*{-0.3cm} &=&\hspace*{-0.3cm}  
  \frac{m_Q^2 (1 + 2 \alpha) - \alpha (1-\alpha ) s}{2 (m_Q^2-\alpha  s)}
  \left(~ 1 \pm\sqrt{1 + \frac{4m_Q^2 \alpha (1-\alpha) (m_Q^2 - \alpha  s)}
  {\Big( m_Q^2(1 + 2\alpha) - \alpha (1 - \alpha) s \Big)^2}} ~\right) 
  \nonumber\\
 \hspace*{-2cm}  \alpha{_{\substack{max \\ min}}}  \hspace*{-0.3cm}&=&  \hspace*{-0.3cm}
  \frac{s - 8m_Q^2}{2s} \left(~ 1 \:\pm\: \sqrt{1 - 
  \frac{4 m_Q^2 s}{(s - 8m_Q^2)^2}} ~\right)  
  \eea
}
while  $\mu_1^n$ and $\mu_2^n$ are defined as\,:
{ \footnotesize
\vspace*{-0.25cm}
\begin{eqnarray}
  {\mu_1^n} &=& \left[m_Q^2 \bigg( \frac{1}{\alpha} + \frac{1}{\beta} + 
  \frac{1}{\gamma} + \frac{1}{1-\alpha-\beta-\gamma}  \bigg)\right]^{\cfrac{n}{2}}  
  \nonumber \\
  {\mu_2^n} &=&\left[ m_Q^2 \bigg( \frac{1}{\alpha} + \frac{1}{\beta} + 
  \frac{1}{\gamma_{max}} + \frac{1}{\gamma_{min}}\bigg)\right]^{\cfrac{n}{2}}
\end{eqnarray}
}
The PT NLO and the LO dimension-8 gluon condensates contributions are estimated using factorization via a convolution of two bilinear vector spectral functions as discussed in Refs.\,\cite{PICH,SNPIVO} and shown literally in Ref.\,\cite{MOLE20}. We assume that the contribution of $\la G^4\ra$ is known within a factor $k_g=(2\pm 1)$. The PT NLO contribution to the vector bilinear two-point function is well-known\,\cite{SNB1,SNB2}. 
\subsection*{\b The Laplace sum rules}
We shall work with the  Finite Energy version of the QCD Inverse Laplace sum rules (LSR) and their ratios\,\cite{BELLa,BELLb,BECCHI,SNR,SNREV}\,:
\bea
 {\cal L}_n(\tau,\mu)&=&\int_{16m_Q^2}^{t_c}\hspace*{-0.5cm}dt~t^n~e^{-t\tau}\frac{1}{\pi} \mbox{Im}~\psi_{X}(t,\mu)~,\nnb\\
 {\cal R}_{n+1,n}(\tau)&=&\frac{{\cal L}_{n+1}} {{\cal L}_n}\,\,: \, \,\, n=0,1. 
\label{eq:lsr}
\eea
 where $m_Q$ is the heavy quark mass, $\tau$ is the LSR variable and $n$ the degree of moments; $t_c$ is  the threshold of the ``QCD continuum" which parametrizes the QCD spectral function  ${\rm Im}\,\psi_{X}(t,m_Q^2,\mu^2)$ above $t_c$.    
\subsection*{\b Minimal Duality Ansatz for the spectral function\label{sec:spectral}}
   In the present case, where no complete data on the spectral function are available, we use the Minimal Duality Ansatz (MDA)\,:
 \bea
\hspace*{-0.7cm}\frac{1}{\pi} {\rm Im} \psi_X(t)\hspace*{-0.3cm}&\simeq&\hspace*{-0.3cm} 2f_X^2 M_X^{8} \delta(t-M_X^2) +\Theta(t-t_c) ``{\rm  QCD Continuum}",
 \eea
 for parametrizing the molecule spectral function. $M_X$ and $f_X$ are the lowest ground state mass and coupling normalized as $f_\pi=93$ MeV.
 Within a such parametrization, one  obtains: 
 \beq
  {\cal R}_{n+1,n}\simeq M_X^2~,
  \eeq
 indicating that the ratio of moments appears to be a useful tool for extracting the mass of the hadron ground state\,\cite{SNB1,SNB2,SNB3,SNB26,SNREV15}. 
\subsection*{\b Optimization criteria\label{sec:optimization}}
\d In order to extract the optimal results, we use stability criteria (plateau, minimum or inflexion point) with respect to the external parameters\,: LSR variable $\tau$,  Continuum threshold $t_c$. The value of the substraction PT scale is fixed to be:
\beq
\mu_c\simeq 4.5\,\rm{GeV}\,\,{\rm and}\,\,\, \mu_b\simeq 7.25\,\rm{GeV}\,
\eeq
from the charm and bottom quark channels\,\cite{HEP18,SU3,QCD16,MOLE16,X5568,MOLE12,MOLE20}. 

\d In addition to these criteria, we shall exclude, in our analysis, the $(\tau,t_c)$ region where the QCD continuum exceeds the Lowest Ground State  Pole contribution to the LSR. It is formulated by the condition\,:
\beq
{\rm R_{P/C}\equiv \frac{ Pole }{QCD\, Continuum}}\geq 1.
\label{eq:rpc}
\eeq
\subsection*{\b The QCD input parameters\label{sec:input}}
We shall use the most recent phenomenological determinations from QCD spectral sum rules,
$e^+e^-\to$ Hadrons and $\tau$-decay data {as shown in Table~\ref{tab:param}}. 

\begin{table}[hbt]
\vspace*{-0.5cm}
\setlength{\tabcolsep}{1.7pc}
    {\scriptsize
\begin{tabular}{lll}
&\\
\hline
Parameters& Values& Refs.    \\
\hline
$\alpha_s(M_\tau)$& $0.3128(51)$&\cite{SNREV25,SNparam}\\
$\overline{m}_c(\overline {m}_c)$&$1286(16)$ MeV&\cite{SNbc20,SNmom18}\\
$\overline{m}_b(\overline {m}_b)$&$4202(8)$ MeV&\cite{SNbc20,SNmom18}\\
$\la\alpha_s G^2\ra$& $(6.35\pm 0.35)\times 10^{-2}$ GeV$^4$&
\cite{SNparam,SNG2}\\
$M_0^2$&$(0.8 \pm 0.2)$ GeV$^2$&\cite{JAMI2,HEID,SNhl}\\
$\la g^3  G^3\ra$& $(8.2\pm 1.0)$ GeV$^2\times\la\alpha_s G^2\ra$&
\cite{SNG3}\\
\hline

\end{tabular}
}
\vspace*{-0.25cm}
    \caption{QCD input parameters. } 
    \label{tab:param}
\end{table}
The value of $\alpha_s$ in Table\,\ref{tab:param} corresponds to\,: 
\bea
&&\hspace*{-1cm}\alpha_s(M_Z)=0.1176(8) \lrar {\hspace*{-0.5cm} \circ} \hspace*{0.35cm}\alpha_s(4.5)=0.2183(29)~{\rm for\,\,} n_f=4\,\nnb\\
&&\hspace*{-1cm}{\rm and}\,\, \alpha_s(7.25)=0.1917(21)~{\rm for\,\,} n_f=5\,{\rm flavours}\,.
\eea
\section{The $2^{++}$ $J/\psi J/\psi$ molecule state}

\begin{figure}[hbt]
\vspace*{-0.25cm}
\begin{center}
\includegraphics[width=7cm]{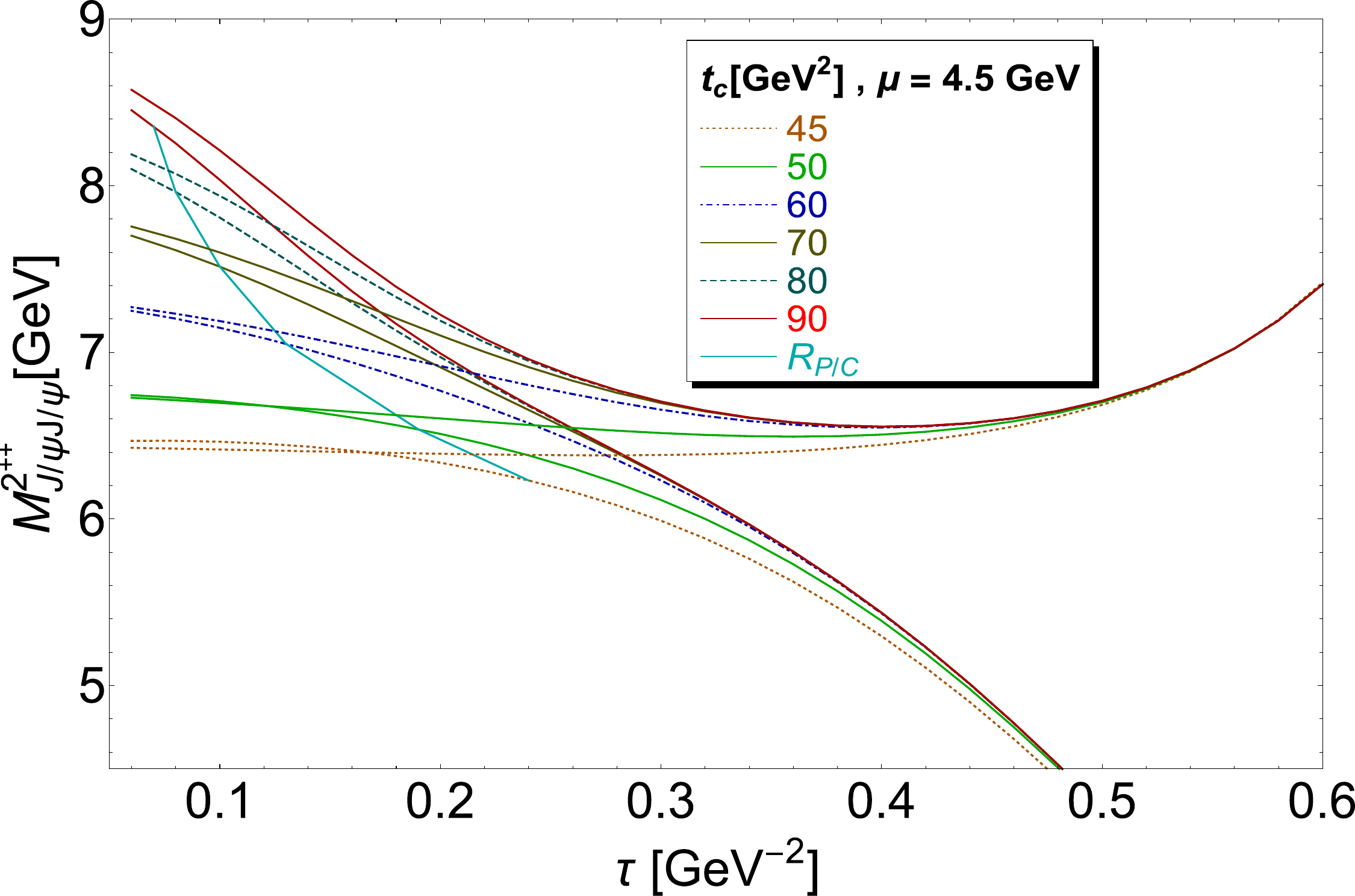}
\vspace*{-0.25cm}
\caption{\footnotesize  $\tau$ and $t_c$ behaviours of the $J/\psi J/\psi$ molecule mass from ${\cal R}_{10}$ (curves with inflexion poins) and ${\cal R}_{21}$ (curves with minimum). The region in the LHS of the $R_{P/C}$ curve is excluded.}
\label{fig:psi-psi-mass}
\end{center}
\vspace*{-0.7cm}
\end{figure} 
\subsection*{\b Lowest ground state mass from ${\cal R}_{10}$}
We use the lowest ratio of moments ${\cal R}_{10}$ by including factorised NLO Pert  and LO $\la G^4\ra$ contributions. The analysis is shown in Fig.\,\ref{fig:psi-psi-mass} (family of curves which decrease when $t_c$ increase). 
We notice that the  $\tau$-stability 
(plateau) for the set $(\tau,t_c)\leq (0.18, 45\sim 50)$\,\footnote{Here and in the following the set  $(\tau,t_c)$ is expressed in units of  (GeV$^{-2}$, GeV$^2$).} is excluded by the constraint $R_{P/C}$ in Eq.\,\ref{eq:rpc}, while the  $t_c$-stability (Ground State Dominance [GSD] of the spectral integral) is reached (inflexion point) for $(\tau,t_c)$=($0.26\pm 0.02$, $\geq 70$). The result is compiled in Table\,\ref{tab:psi-psi-2}.    
\subsection*{\d Lowest ground state mass from ${\cal R}_{21}$}
We do the same analysis for the ratio of moments ${\cal R}_{21}$. The result is shown in Fig.\,\ref{fig:psi-psi-mass}.
(family of curves with minimum). We obtain, for the set of $(\tau,t_c)$=(0.44, 45)  (beginning of $\tau$-stability) and $(\tau,t_c)$=(0.5, 60) (beginning of $t_c$-stability),\,
the results given in Table\,\ref{tab:psi-psi-2}.
\subsection*{\b Final value of the lowest ground state mass}
We consider as a final {\it conservative result} the weighted average of results from the two moments ${\cal R}_{10}$ and ${\cal R}_{21}$ shown in Table\,\ref{tab:psi-psi-2}.\  
\subsection*{\b Lowest ground state coupling from ${\cal L}_{0}$ }
\begin{figure}[hbt]
\vspace*{-0.25cm}
\begin{center}
\includegraphics[width=7cm]{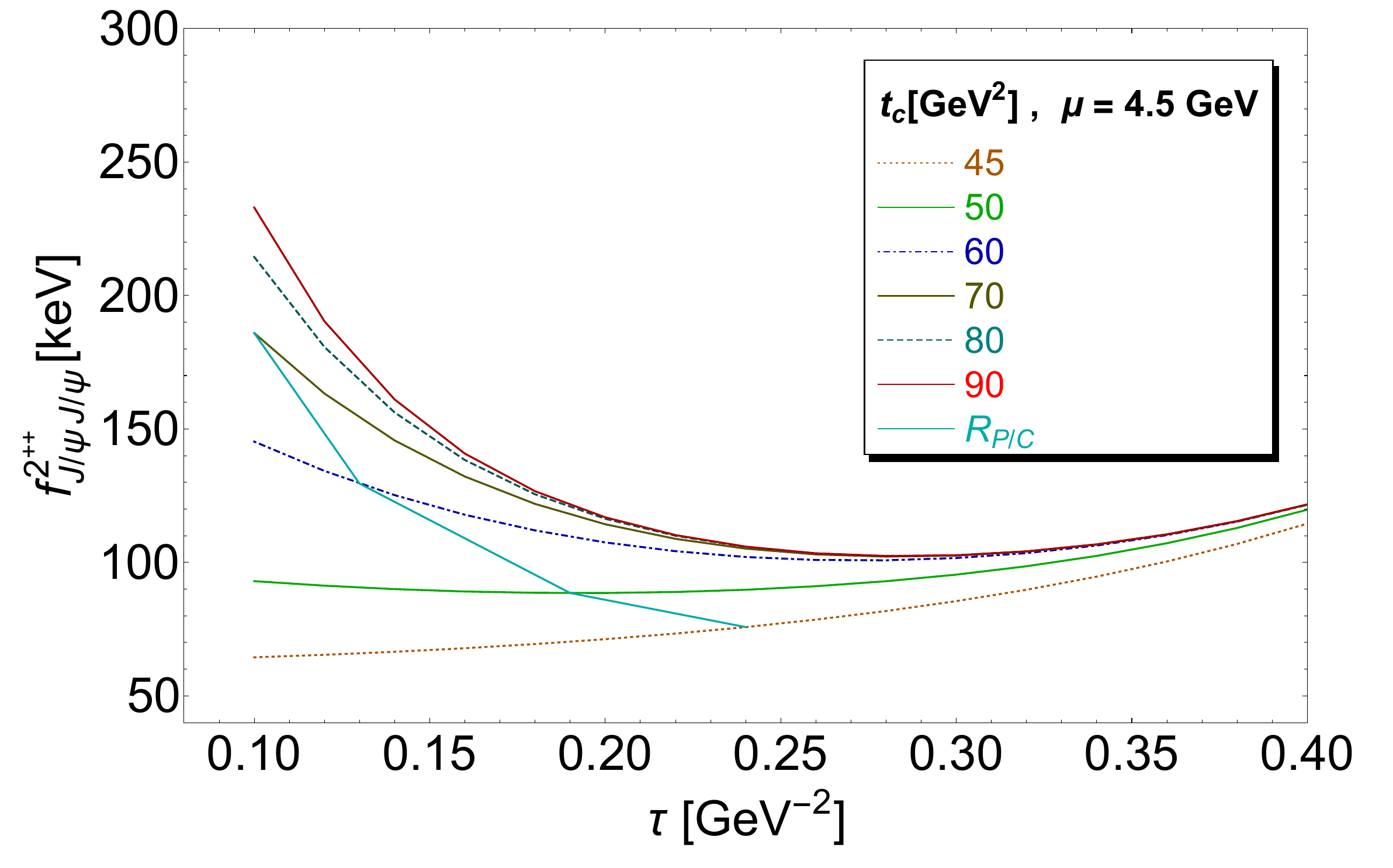}
\vspace*{-0.25cm}
\caption{\footnotesize  $\tau$ and $t_c$ behaviours of the $J/\psi J/\psi$ molecule coupling from ${\cal L}_{0}$. The region in the LHS of the $R_{P/C}$ curve is excluded.}
\label{fig:psi-psi-coupling}
\end{center}
\vspace*{-0.5cm}
\end{figure} 
 
Using the previous value of the mass as input in the ${\cal L}_0$ moment, we show the $\tau$ and $t_c$ behaviours of the coupling in Fig.\,\ref{fig:psi-psi-coupling}.   We obtain the mean value from the sets $(\tau,t_c) = (0.20, 50)$ and (0.28,$\geq$ 80) which is given in Table\,\ref{tab:psi-psi-2}.
\begin{figure}[H]
\vspace*{-0.25cm}
\begin{center}
\includegraphics[width=7cm]{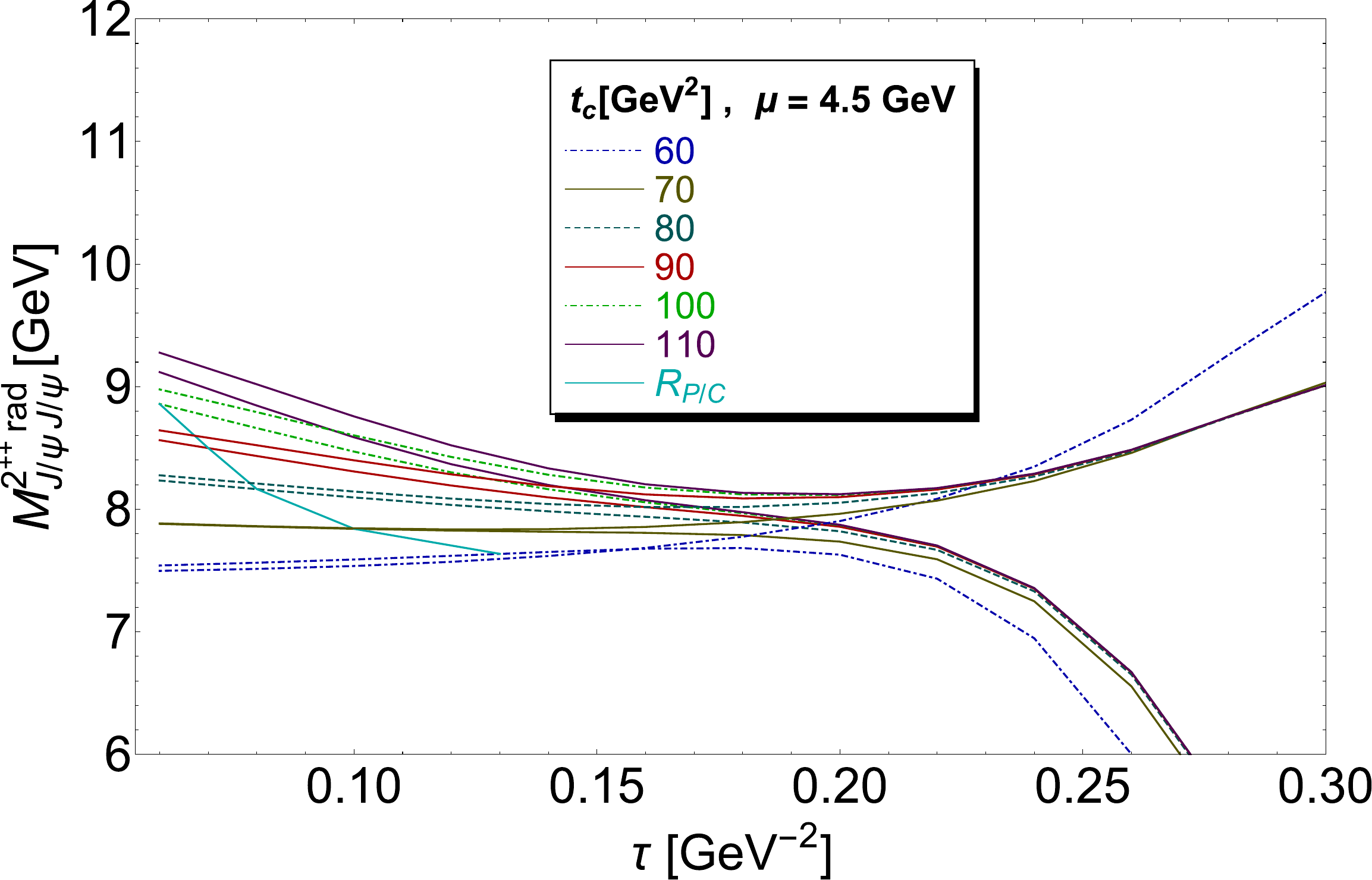}
\vspace*{-0.25cm}
\caption{\footnotesize  $\tau$ and $t_c$ behaviours of the $J/\psi J/\psi$ molecule first radial excitation mass from ${\cal R}_{10}$ (curves with inflexion points) and ${\cal R}_{21}$ (curves with minimum). The region in the LHS of the $R_{P/C}$ curve is excluded.}
\label{fig:psi-psi-mass-rad-2}
\end{center}
\vspace*{-0.25cm}
\end{figure} 
\subsection*{\b First radial excitation mass}
We attempt to extract the 1st radial excitation mass using a ``two-resonances $\oplus$ QCD continuum" parametrization of the spectral function and as input the previous values of the mass and coupling of the lowest ground state obtained previously. The analysis is shown in Fig.\,\ref{fig:psi-psi-mass-rad-2}. 

 \d Using ${\cal R}_{10}$, we obtain optimal results $M^{2^{++}\rm{rad}}_{J/\psi  J/\psi}$ = 7826 and 8163 MeV for the sets $(\tau,t_c)$ =(0.12,70) and ($0.14\pm 0.01$, 100) from which we deduce the results in Table\,\ref{tab:psi-psi-rad-2}.

\d  For ${\cal R}_{21}$, the optimal results :  $M^{2^{++}\rm{rad}}_{J/\psi  J/\psi}$ = 7837 and 8117 MeV are obtained from the sets $(\tau,t_c)$ = (0.12,70) and (0.20,100), from which we deduce the results in Table\,\ref{tab:psi-psi-rad-2}.

 \d The final value is the weighted average of the two previous predictions quoted in  Table\,\ref{tab:psi-psi-rad-2}.
\begin{table*}[hbt]
\vspace*{-1.cm}
\setlength{\tabcolsep}{0.3pc}
    {\scriptsize
\begin{tabular*}{\textwidth}{@{}l@{\extracolsep{\fill}}lllllllclll ll}
&\\
\hline
\it  \bm $2^{++}\,Ground\, State$&$ t_c$&$\tau$& $\alpha_s$& $m_c$&$G^2$&$G^3$&$G^4$&$ M_X$ [MeV]&$\Delta M_X$&$f_X$ [keV] \\
\hline
\oliva{\it Molecule}\\  
${\cal R}_{10}$& 5&136&11& 29&23&36&27&6526(148)\\
${\cal R}_{21}$&34&6&17& 41&26&31&35&6520(78) \\
\it Mean &&&&&&&&\it \oliva 6521(69) \\
${\cal L}_{0}$&6.8&0.7&3.1&6.9&0.7&0.1&0.7&--&8.7&\oliva 96(14) \\   

\purple\it Four-quark \\
${\cal R}_{10}$& 190&105&1&18&21&26&21&6480(221)\\
${\cal R}_{21}$&87&2.5&21&48&17&18&2&6314(111) \\
\it Mean &&&&&&&&\it \purple 6347(99) \\
${\cal L}_{0}$& 2.6&1.&6.3&13.5&0.2&0.3&0.5&--&24.6&\purple 151(29)\\

\hline
\boldmath  $2^{++}\, 1st\, Radial$ &$ t_c$&$\tau$& $\alpha_s$& $m_c$&$G^2$&$G^3$&$G^4$&$M_X$&$f_X$&$ M_{X'}$ [MeV]&$\Delta_{M_{X'}}$&$f_{X'}$ [keV] \\
\hline
\oliva{\it Molecule}\\
${\cal R}_{10}$&169&123&39&95
&7&2&27&83&226& 7995(336)\\
${\cal R}_{21}$&140&29&
66&158&20&10&68&269&187&7977(378) \\
\it Mean &&&&&&&&&&\it\oliva 7987(251) \\
${\cal L}_{0}$&4.4&1.2&25.8 &9.2 &3.2&0.3 &24.1  &11.1&22&--&53.8&\oliva 248(70) \\  
 
\purple\it Four-quark \\
${\cal R}_{10}$& 68&14&101&51&1&8&5&118&237&7731(272)\\
${\cal R}_{21}$&82&28&79&173&6&3&7&238&208&7755(379) \\
\it  Mean &&&&&&&&&&\it \purple 7739(221) \\
${\cal L}_{0}$&21 &7&10&28&6&5&7&15&46&--&75&\purple 453(96)\\

\hline

\it \boldmath $0^{++}\, Ground\, State$&$ t_c$&$\tau$& $\alpha_s$& $m_c$&$G^2$&$G^3$&$G^4$&$ M_X$ [MeV]&$\Delta M_X$&$f_X$ [keV] \\
\hline
\oliva{\it Molecule}\\  
${\cal R}_{10}$& 111&89&18& 37&1&11&17&6365(149.5)\\
${\cal R}_{21}$&147&91& 17&35&3&7&18&6434(178.3) \\
\it Mean &&&&&&&&\it \oliva 6394(115) \\
${\cal L}_{0}$&8&1&7&15&2&1&8&--&33&\oliva 194(39) \\   

\purple\it Four-quark \\
${\cal R}_{10}$&103&92&20&39&4&11&0.1&6351(145.4)\\
${\cal R}_{21}$&141&95&19&38&2&7&1&6423(175.4) \\
\it Mean &&&&&&&&\it \purple 6380(112) \\
${\cal L}_{0}$&10&4&11&25&1&2&1&--&49&\purple 303(57)\\

\hline
\it\boldmath $0^{++}\,1st\, Radial$ &$ t_c$&$\tau$& $\alpha_s$& $m_c$&$G^2$&$G^3$&$G^4$&$M_X$&$f_X$&$ M_{X'}$ [MeV]&$\Delta_{M_{X'}}$&$f_{X'}$ [keV] \\
\hline
\oliva{\it Molecule}\\
${\cal R}_{10}$&91&16&47&78&17&9&2&149&367& 7815(417)\\
${\cal R}_{21}$&89&18&44&81&12&6&13&127&347&7807(392) \\
\it Mean &&&&&&&&&&\it\oliva 7811(286) \\

${\cal L}_{0}$&37&11&21 &27&15&10&18&15&37&--&86&\oliva 485(111) \\  
 
\purple\it Four-quark \\
${\cal R}_{10}$& 108&19&58&98&6&3&6&126&330&7853(387)\\
${\cal R}_{21}$&104&20&51&96&12&12&9&113&319&7840(371) \\
\it  Mean &&&&&&&&&&\it \purple 7846(268) \\
${\cal L}_{0}$&38 &21 & 33&39&13&9&17&23&57&--&127&\purple 769(158)\\

\hline

\end{tabular*}
}
\vspace*{-0.25cm}
      \caption{{\it\bf Charm quark channel}\,: Masses from the ratios of moments ${\cal R}_{10}$ and ${\cal R}_{21}$ and couplings from the moment ${\cal L}_0$ of the lowest ground states and first radial excitations. The ``mean" is the weighted average from ${\cal R}_{10}$ and ${\cal R}_{21}$.The different sources of errors are  in units of MeV for the masses and in keV for the couplings.} 
     
       \label{tab:psi-psi-2}   \label{tab:psi-psi-rad-2}
    \label{tab:psi-psi-0}
\end{table*}


\subsection*{\b First radial excitation coupling}
\begin{figure}[H]
\vspace*{-0.25cm}
\begin{center}
\includegraphics[width=7cm]{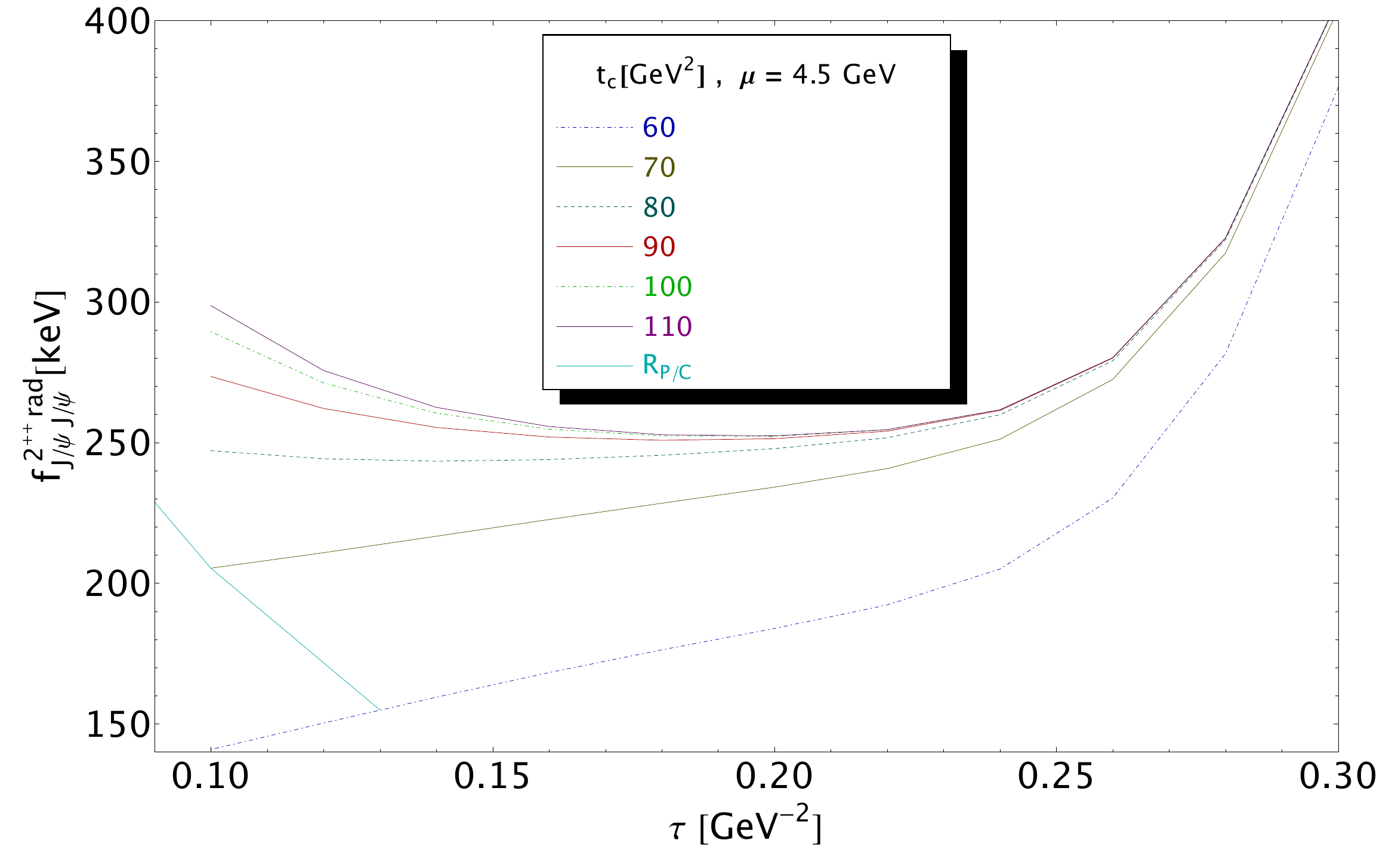}
\vspace*{-0.25cm}
\caption{\footnotesize  $\tau$ and $t_c$ behaviours of the $J/\psi J/\psi$ molecule first radial excitation coupling from ${\cal L}_{0}$. The region in the LHS of the $R_{P/C}$ curve is excluded.}
\label{fig:psi-psi-coupling-rad-2}
\end{center}
\vspace*{-0.5cm}
\end{figure}

We use ${\cal L}_0$ to extract the coupling.  The result of the analysis is shown in Fig.\,\ref{fig:psi-psi-coupling-rad-2}. 

We obtain for  $(\tau,t_c)$  from (0.14,80) to (0.2,100) the value 243.4 to 252.2 keV 
which leads to the one in Table\,\ref{tab:psi-psi-rad-2}.

\section{The $2^{++}$ ${cc\bar c\bar c}$ four-quark states}
We estimate the mass using the ratio of LSR moments ${\cal R}_{10}$ and  ${\cal R}_{21}$. 
The analysis  is similar to the one of the molecule where the curves versus $(\tau,t_c)$ show the same behaviours. 

\d For ${\cal R}_{10}$, we obtain for the set $(\tau,t_c)=(0.24, 45)$  (beginning of $\tau$-stability) allowed by the $R_{P/C}$ condition\,: $M^{2^{++}}_{cc\bar c\bar c}\simeq 6290 \,{\rm MeV}$.  The $t_c$-stability (GSD of the spectral integral) is reached (inflexion point) for $(\tau,t_c)$=($0.26\pm 0.02, \geq 70$) at which $
M^{2^{++}}_{cc\bar c\bar c}\simeq 6669 \,{\rm MeV}$.  The results are given in Table\,\ref{tab:psi-psi-2}.

\d We do a similar analysis for the ${\cal R}_{21}$ moment. 
At the $\tau$-minimum, we obtain $M^{2^{++}}_{cc\bar c\bar c}=6271$ MeV for the set $(\tau,t_c) = (0.44,45)$  (beginning of $\tau$-stability) and $M^{2^{++}}_{cc\bar c\bar c}=6358$ MeV for the set $(\tau,t_c) = (0.5,60)$ (beginning of $t_c$-stability) from which we deduce the value given in Table\,\ref{tab:psi-psi-2}.
\d We consider as a final value of the ground state mass the weighted average of the two determinations quoted in Table\,\ref{tab:psi-psi-2}.

\subsection*{\b The lowest ground state coupling from ${\cal L}_{0}$ }

The coupling is better determined from the moment ${\cal L}_0$. The behaviour of the curves with respect to the variation of the set $(\tau,t_c)$ is the same as for the molecule.  The $\tau$-minimum starts for  $(\tau=0.2, t_c=50)$ and a $t_c$-stablity for $(\tau=0.3, t_c=80)$.  We deduce in this range of $t_c$-values the value quoted in Table\,\ref{tab:psi-psi-2}.

\subsection*{\b First radial excitation mass}
Doing an  analysis similar to the molecule state, we obtain curves similar to the ones in Fig.\,\ref{fig:psi-psi-mass-rad-2}. 

\d For ${\cal R}_{10}$, we deduce from the sets  $(\tau,t_c)$ = (0.16,70) and (0.22,100) the values \,:$M^{2^{++}{\rm rad}}_{cc\bar c\bar c}$= 7663 and 7792 MeV leading to the result quoted in Table\,\ref{tab:psi-psi-rad-2}.

\d For ${\cal R}_{21}$, we deduce from the sets  $(\tau,t_c)$ = (0.18,70) and (0.22,100) the values \,:$M^{2^{++}{\rm rad}}_{cc\bar c\bar c}$= 7673 and 7837 MeV leading to the value quoted in Table\,\ref{tab:psi-psi-rad-2}.

\d The two previous determinations lead to the weighted average quoted in Table\,\ref{tab:psi-psi-rad-2}. 
\vspace*{-0.7cm}
\subsection*{\b First radial excitation coupling}
\begin{figure}[hbt]
\vspace*{-0.25cm}
\begin{center}
\includegraphics[width=7cm]{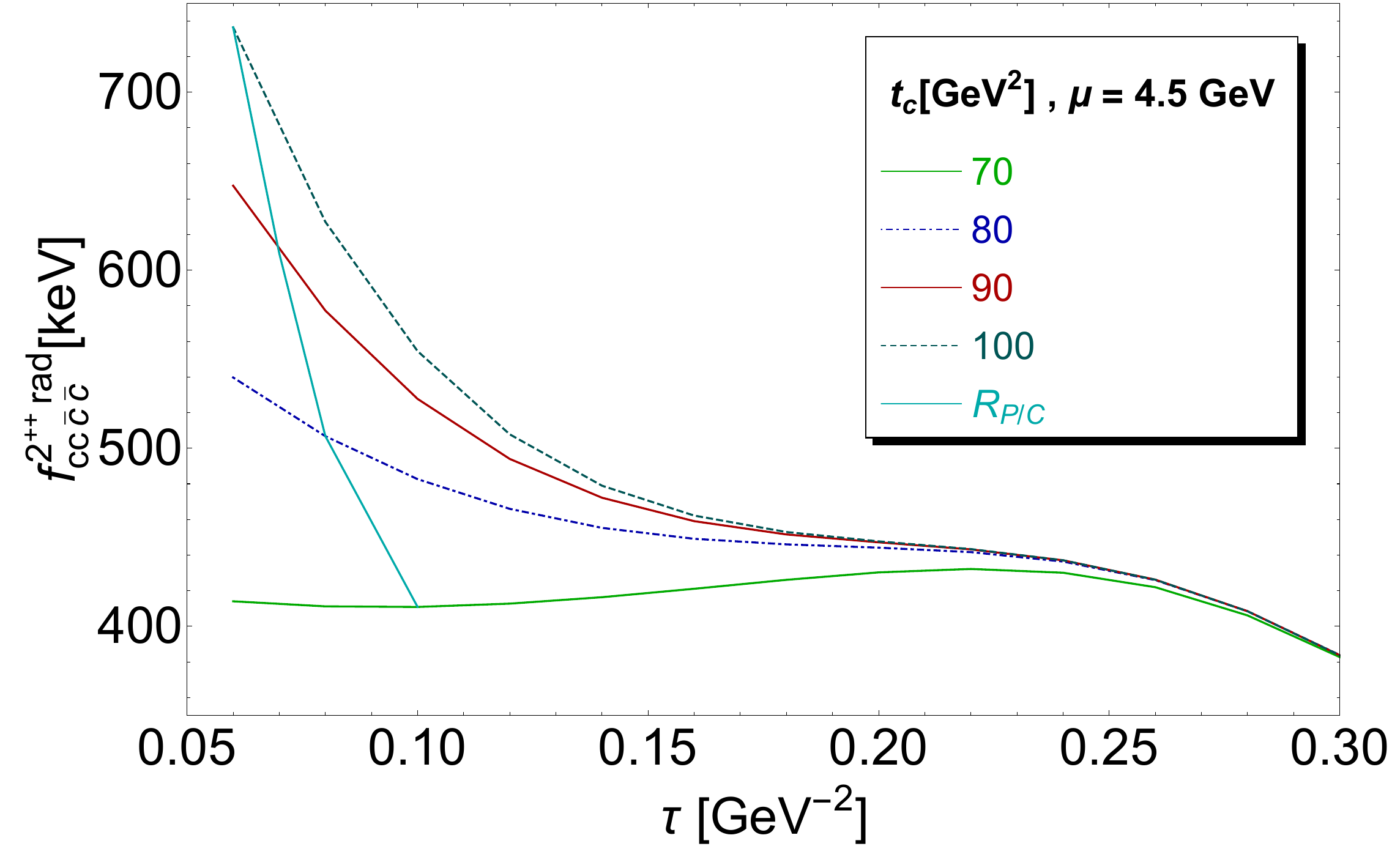}
\vspace*{-0.25cm}
\caption{\footnotesize  $\tau$ and $t_c$ behaviours of the four-quark first radial excitation coupling from ${\cal L}_{0}$. The region in the LHS of the $R_{P/C}$ curve is excluded.}
\label{fig:4c-coupling-rad-2}
\end{center}
\vspace*{-0.5cm}
\end{figure}

We extract the coupling from ${\cal L}_0$. The analysis is shown in Fig.\,\ref{fig:4c-coupling-rad-2}. Our results quoted in Table\,\ref{tab:psi-psi-rad-2} come from the range $(\tau,t_c)$ = (0.1,70) to (0.19,100). 


\section{The $0^{++}\,J/\psi J/\psi$ molecule state}
In this section, we improve our previous predictions obtained in Ref.\,\cite{MOLE20} by including the contribution of the $\la G^4\ra$ condensates roughly estimated in Ref.\,\cite{MOLE20} and by considering the results from the ratio of moments ${\cal R}_{21}$. The QCD expression of the corresponding spectral function is given explicitly in Ref.\,\cite{MOLE20}.
\subsection*{\b The lowest ground state mass }
\begin{figure}[hbt]
\vspace*{-0.25cm}
\begin{center}
\includegraphics[width=7cm]{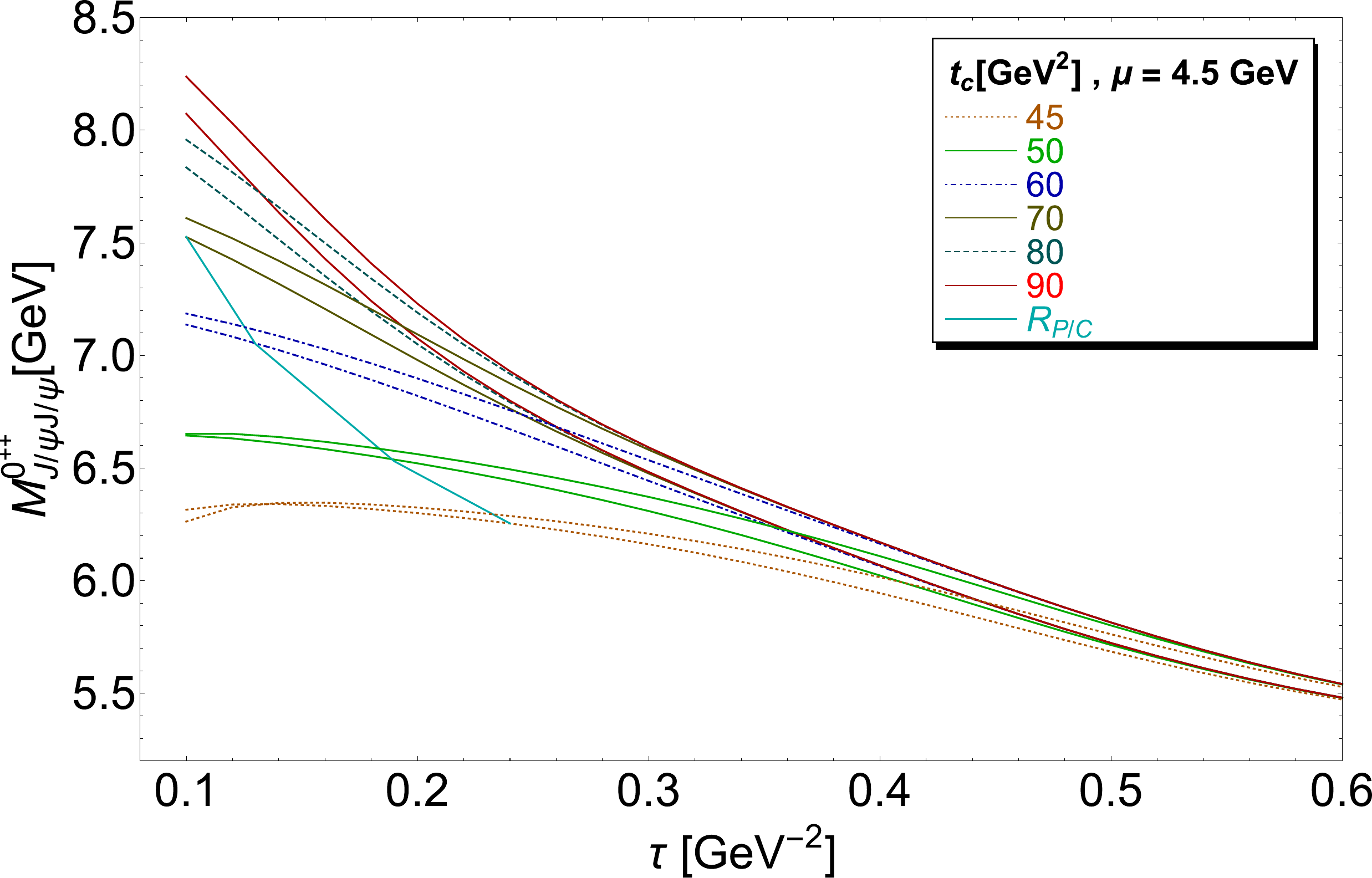}
\vspace*{-0.25cm}
\caption{\footnotesize  The same as in Fig.\,\ref{fig:psi-psi-mass} but for the $0^{++}$ scalar molecule  mass. The low family of curves is from ${\cal R}_{10}$ while the high one is from ${\cal R}_{21}$.}
\label{fig:mass-psipsi-0}
\end{center}
\vspace*{-0.5cm}
\end{figure}

The analysis is similar to the one in the previous section. It is shown in Fig.\,\ref{fig:mass-psipsi-0} (low family of curves with inflexion point). 
 
\d  For ${\cal R}_{10}$, $\tau$ stability starts for the set $(\tau,t_c)=(0.1\sim 0.24, 45)$ while the $t_c$-stability (GSD of the spectral integral) is reached (inflexion point) for $(\tau,t_c)$=($0.30\pm 0.02, \geq 70$) at which we deduce respectively:
$M^{0^{++}}_{J/\psi J/\psi}\simeq 6254~{\rm and}~6476 \,{\rm MeV}\, $ and the mean given in Table\,\ref{tab:psi-psi-0}. 

\d A similar analysis is done for ${\cal R}_{21}$ and shown in Fig.\,\ref{fig:mass-psipsi-0} (higher family of curves). We obtain for the sets $(\tau,t_c)$ = (0.24,45) and (0.30,70) the values\,: $M^{0^{++}}_{J/\psi J/\psi}\simeq 6287~{\rm and}~6581 \,{\rm MeV}\,$ from which we deduce the mean given in Table\,\ref{tab:psi-psi-0}. 

\d We consider as a final value of the ground state mass the weighted average of the two determinations given in Table\,\ref{tab:psi-psi-0}. 

\subsection*{\b The molecule lowest ground state coupling from ${\cal L}_{0}$ }
\begin{figure}[hbt]
\vspace*{-0.25cm}
\begin{center}
\includegraphics[width=7cm]{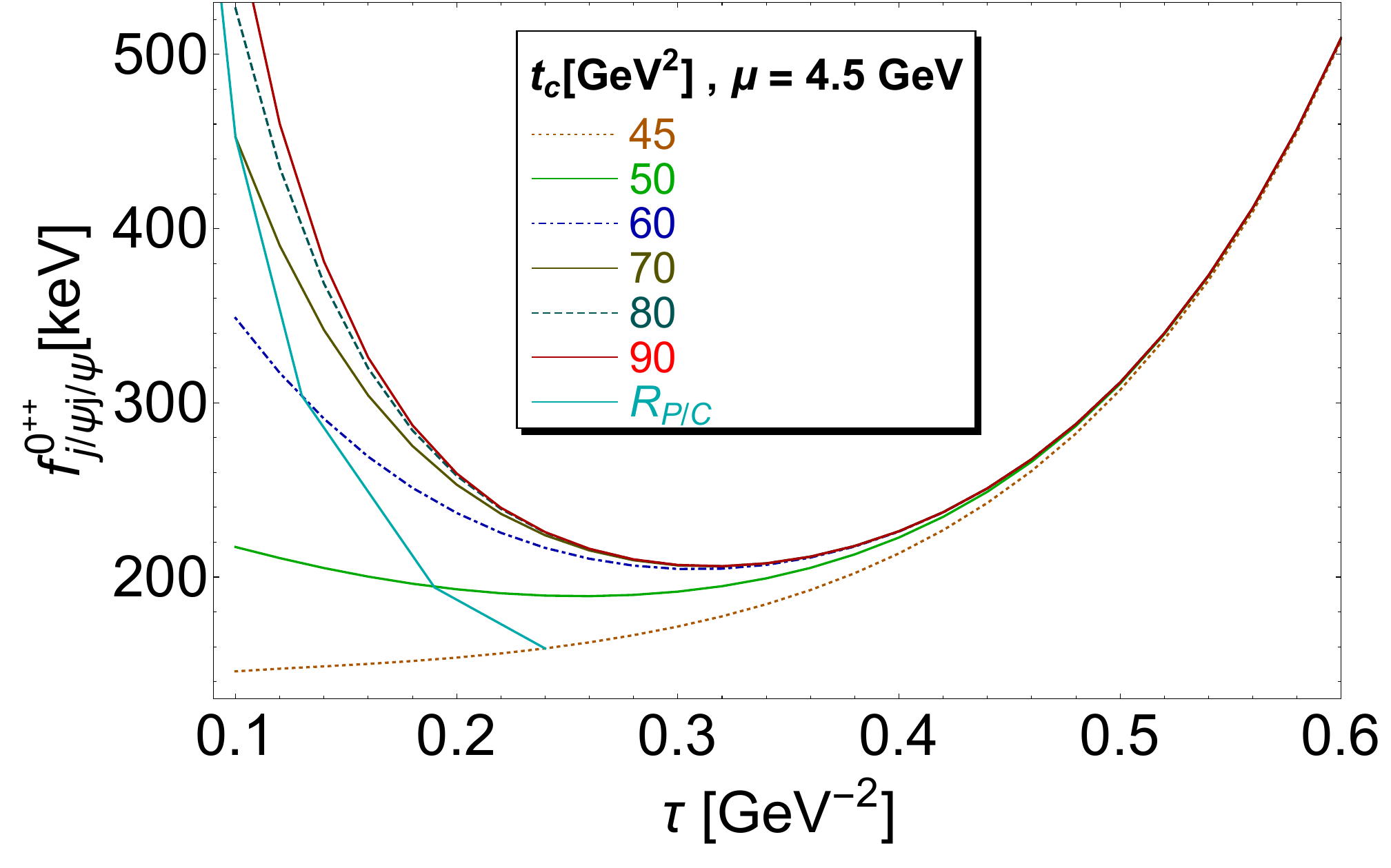}
\vspace*{-0.25cm}
\caption{\footnotesize  The same as in Fig.\,\ref{fig:psi-psi-coupling} but for the $0^{++}$ scalar molecule  coupling.}
\label{fig:coupling-psipsi-0}
\end{center}
\vspace*{-0.5cm}
\end{figure}

The analysis is shown in Fig.\,\ref{fig:coupling-psipsi-0}. One can notice that there is a minimum in $\tau$ from the set $(\tau,t_c)\simeq (0.26, 50)$  to the $t_c$-stability (0.32, 80) to which correspond the values 293.3 and 313.1 keV. Their mean value is given in Table\,\ref{tab:psi-psi-0}. 
\subsection*{\b The molecule first radial excitation }
\d The mass is extracted from ${\cal R}_{10}$ for $(\tau,t_c)$  from (0.14,70)\,: $M^{0^{++}\rm{rad}}_{J/\psi J/\psi}$=7724 MeV  to (0.20,100)\,: $M^{0^{++}\rm{rad}}_{J/\psi J/\psi}$=7906 MeV and from ${\cal R}_{21}$ from (0.16,70)\,: $M^{0^{++}\rm{rad}}_{J/\psi J/\psi}$=7717 MeV  to (0.22,100)\,: $M^{0^{++}\rm{rad}}_{J/\psi J/\psi}$=7897 MeV leading to the results given in Table\,\ref{tab:psi-psi-0}. 

\d The coupling is extracted from ${\cal L}_{0}$.  $(\tau,t_c)$ stabilities are obtained for (0.23,60) to (0.28,100) corresponding to  $f^{0^{++}\rm{rad}}_{J/\psi J/\psi}$ = 448 to 521 keV. Final result is givenin Table\,\ref{tab:psi-psi-0}.

\section{The $0^{++}$ ${cc\bar c\bar c}$ four-quark state}
\subsection*{\b The lowest ground state mass}
The behaviours of the curves for ${\cal R}_{10}$ and ${\cal R}_{21}$ are similar to the one for the molecule shown in Fig.\,\ref{fig:mass-psipsi-0}. For  ${\cal R}_{10}$,  $\tau$-stability starts from the set $(\tau,t_c)$ = (0.24, 45) while the $t_c$-stability is reached for the set $(\tau,t_c)$ = (0.3, 70)  to which correspond\,: $
M^{0^{++}}_{cc\bar c\bar c}\simeq 6248~  {\rm and} ~ 6454\,{\rm MeV},$ and the mean given in Table\,\ref{tab:psi-psi-0}.  
For  ${\cal R}_{21}$, the  mean in Table\,\ref{tab:psi-psi-0} comes from the set (0.24,45) and (0.30,70) corresponding to $
M^{0^{++}}_{cc\bar c\bar c}\simeq 6282~  {\rm and} ~ 6563\,{\rm MeV},$

\subsection*{\b The lowest ground state coupling from ${\cal L}_{0}$ }
The behaviour of the curves is similar to the case of the $J/\psi J/\psi$ molecule shown in Fig.\,\ref{fig:coupling-psipsi-0}. $\tau$-stability starts from the set $(\tau,t_c)$ = (0.26,50) while $t_c$ stability is reached for $(\tau,t_c)$ = (0.32,80) to which corresponds respectively the value 293 and 313 keV from which we deduce the value quoted in Table\,\ref{tab:psi-psi-0}.

\subsection*{\b The $cc\bar c\bar c$ first radial excitation }

\d The mass is extracted from ${\cal R}_{10}$ for $(\tau,t_c)$  from (0.14,70)\,: $M^{0^{++}\rm{rad}}_{{cc\bar c\bar c}}$=7744 MeV  to (0.20,100)\,: $M^{0^{++}\rm{rad}}_{{cc\bar c\bar c}}$=7959 MeV and from ${\cal R}_{21}$ from (0.16,70)\,: $M^{0^{++}\rm{rad}}_{{cc\bar c\bar c}}$=7736 MeV  to (0.22,100)\,: $M^{0^{++}\rm{rad}}_{cc\bar c\bar c}$=7944 MeV leading to the results quoted in Table\,\ref{tab:psi-psi-0}. The behaviour of the curves is similar to the case of molecule.

\d The coupling is extracted from ${\cal L}_{0}$.  $(\tau,t_c)$ stabilities are obtained for (0.23,60) to (0.28,100) corresponding to  $f^{0^{++}\rm{rad}}_{cc\bar c\bar c}$ = 731 to 807 keV. Final result is givenin Table\,\ref{tab:psi-psi-0}.

\section{Extension of the analysis to the $\Upsilon\Upsilon$ family}
\begin{figure}[hbt]
\vspace*{-0.2cm}
\begin{center}
\includegraphics[width=7cm]{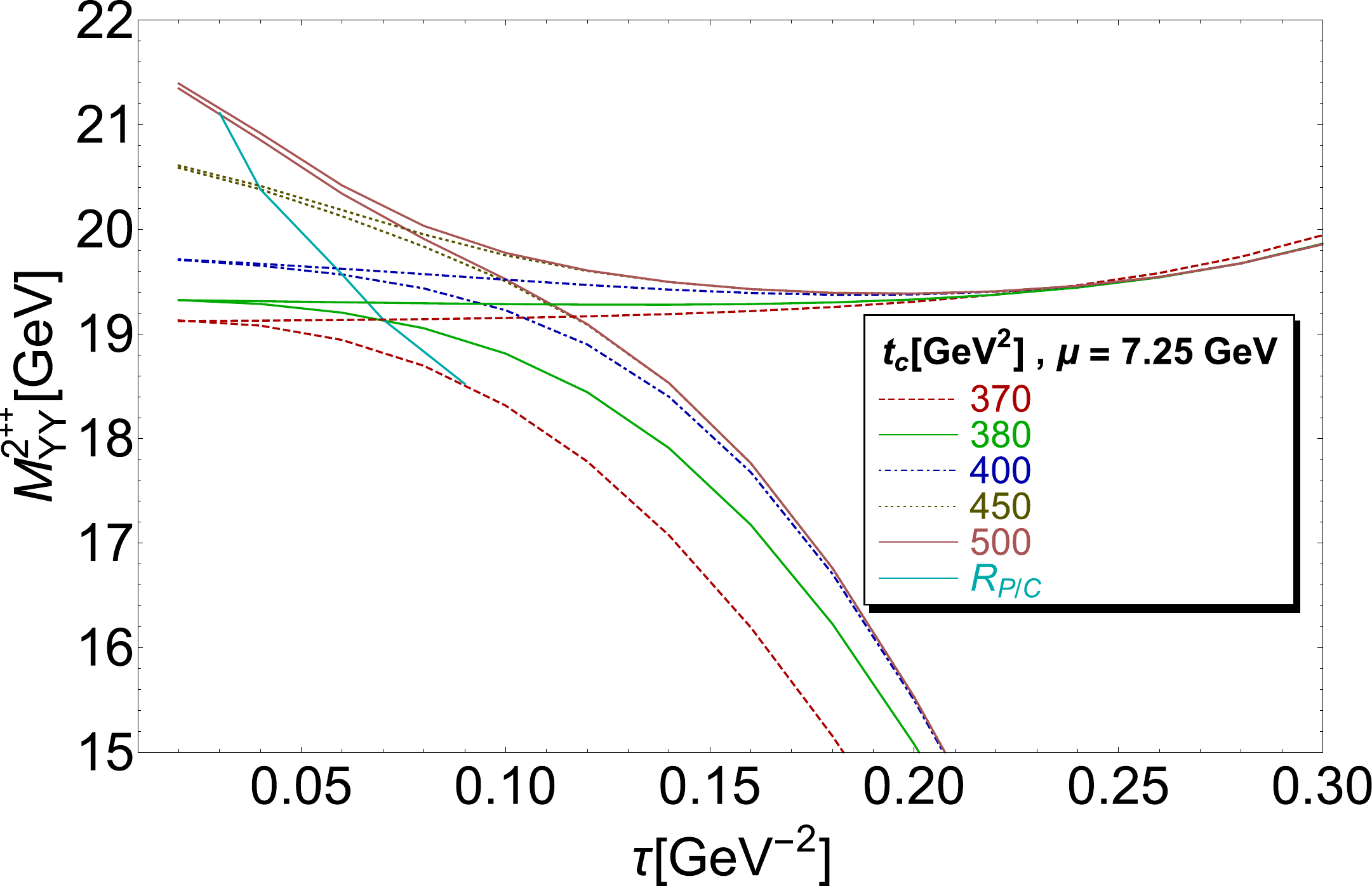}
\vspace*{-0.25cm}
\caption{\footnotesize  The same as in Fig.\,\ref{fig:psi-psi-mass} but for  the $2^{++}\,\Upsilon\Upsilon$ molecule  mass.}
\label{fig:massb-2}
\end{center}
\vspace*{-0.5cm}
\end{figure}

The strategy is similar to the charm quark channel after changing the scale to $\mu_b=7.25$ GeV at which are evaluated the QCD parameters. We illustrate the analysis by the determination of the $2^{++}\, \Upsilon\Upsilon$ molecule mass  shown in Fig.\,\ref{fig:massb-2}.
\subsection*{\b The $2^{++}\,\Upsilon\Upsilon$ molecule groung state mass and coupling}
 \d  ${\cal R}_{10}$ is (almost) $\tau$-stable for $t_c\simeq$ 380 GeV$^2$ where $M^{2^{++}}_{\Upsilon\Upsilon}\simeq$19139 MeV. The inflexion point starts from $(\tau,t_c)$ = (0.10,450) where  $M^{2^{++}}_{\Upsilon\Upsilon}$=19503 MeV. Their mean is given in Table\,\ref{tab:upsilon-2}.     
 
 \d For  ${\cal R}_{21}$, the $\tau$ stability  starts from (0.14,380) while $t_c$-stability starts from (0.20,450) where $M^{2^{++}}_{\Upsilon\Upsilon}$= 19280 and 19386. Their mean and the final estimate are given in Table\,\ref{tab:upsilon-2}. 

\begin{figure}[hbt]
\vspace*{-0.25cm}
\begin{center}
\includegraphics[width=7cm]{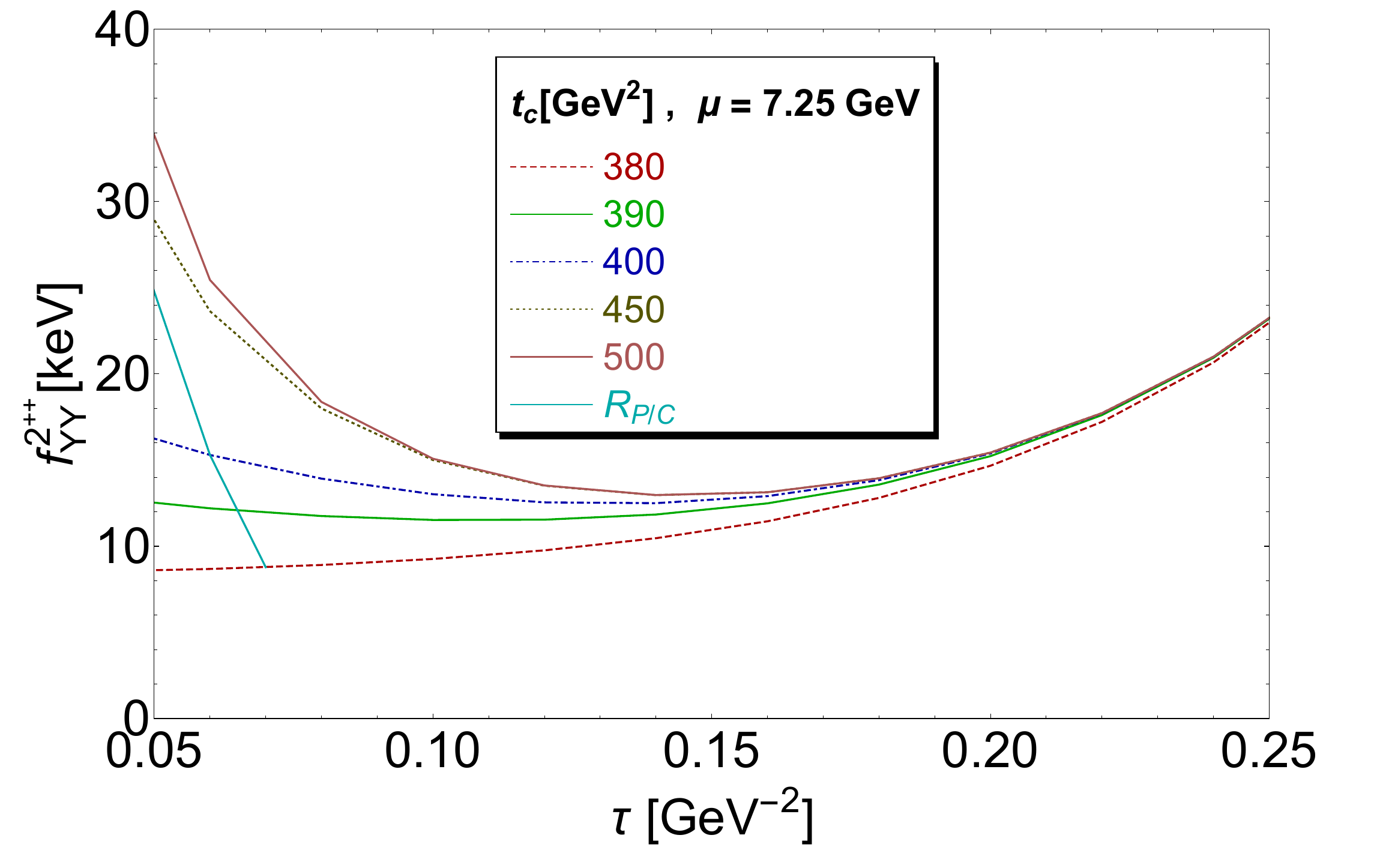}
\vspace*{-0.25cm}
\caption{\footnotesize  The same as in Fig.\,\ref{fig:psi-psi-coupling} but for the $2^{++} \Upsilon\,\Upsilon$ molecule  coupling.}
\label{fig:Upsilon-coupling-2}
\end{center}
\vspace*{-0.75cm}
\end{figure}

\d The analysis of the coupling is shown in Fig.\,\ref{fig:Upsilon-coupling-2}.   We deduce from $(\tau,t_c)$= (0.10,390) to ($0.14$, 500) the values 11.5 and 13 keV leading to the mean given in Table\,\ref{tab:upsilon-2}. 
\begin{table*}[hbt]
\vspace*{-0.5cm}
\setlength{\tabcolsep}{0.3pc}
    {\scriptsize
\begin{tabular*}{\textwidth}{@{}l@{\extracolsep{\fill}}lllllll l  lll ll}
&\\
\hline
\it \boldmath $2^{++}\,Ground\, State$&$ t_c$&$\tau$& $\alpha_s$& $m_b$&$G^2$&$G^3$&$G^4$&$ M_X$ [MeV]&$\Delta M_X$&$f_X$ [keV] \\
\hline
\oliva{\it Molecule}\\  
${\cal R}_{10}$& 182&183&29& 31&11&26&0&19321(263)\\
${\cal R}_{21}$&54&4&31&36&11&19&0&19333(75) \\
\it Mean &&&&&&&&\it \oliva 19332(72) \\
${\cal L}_{0}$&0.7&0.4&1.8&1.4&0.0&0.1&0.0&--&2.4&\oliva 12.3(3.4) \\   

\purple\it Four-quark \\
${\cal R}_{10}$&71&128&31&33&5&13&0&19613(155)\\
${\cal R}_{21}$&26&5&31&36&10&12&0&19264(57) \\
\it Mean &&&&&&&&\it \purple 19306(54) \\
${\cal L}_{0}$& 0.4&0.4&3.1&2.6&0.0&0.1&0.0&--&3.2&\purple 20(5)\\

\hline	
\it\boldmath $2^{++}\, 1st\, Radial $&$ t_c$&$\tau$& $\alpha_s$& $m_b$&$G^2$&$G^3$&$G^4$&$M_X$&$f_X$&$ M_{X'}$ [MeV]&$\Delta_{M_{X'}}$&$f_{X'}$ [keV] \\
\hline
\oliva{\it Molecule}\\
${\cal R}_{10}$&353&204&64&75
&1&4&0&75&232& 20350(485)\\
${\cal R}_{21}$&113&33&
152&169&3&3&3&237&451&20410(570) \\
\it Mean &&&&&&&&&&\it\oliva 20375(369) \\
${\cal L}_{0}$&0.5&1&5 &6 &0&1 &0 &4&7&--&17&\oliva 54(20) \\  
 
\purple\it Four-quark \\
${\cal R}_{10}$&237&103&117&137&1.3&5.3&0.1&109&374&20184(501)\\
${\cal R}_{21}$&89&26&101& 117& 0&1&0&174&394&20343(467) \\
\it  Mean &&&&&&&&&&\it \purple 20269(342) \\
${\cal L}_{0}$&2.6 &3 &9 &11&0&3&0&7&13&--&24&\purple 89(32)\\

\hline
\it \boldmath $0^{++}\,Ground\, State$&$ t_c$&$\tau$& $\alpha_s$& $m_b$&$G^2$&$G^3$&$G^4$&$ M_X$ [MeV]&$\Delta M_X$&$f_X$ [keV] \\
\hline
\oliva{\it Molecule}\\  
${\cal R}_{10}$&145&78&31&34&1&4&0&19296(171)\\
${\cal R}_{21}$&151&80&30&33&1&4&0&19309(177) \\
\it Mean &&&&&&&&\it \oliva  19302(123) \\
${\cal L}_{0}$&3&1&3&3&0&0.2&0&--&9&\oliva 26(10) \\   

\purple\it Four-quark \\
${\cal R}_{10}$&178&85&31&34&2&6&0&19348(203)\\
${\cal R}_{21}$&186&85&30&33&2&5&0&19364(209) \\
\it Mean &&&&&&&&\it \purple  19356(146) \\
${\cal L}_{0}$&4&1.4&3&4&0.1&0.3&0&--&15&\purple 41(16)\\
\hline	
\it\boldmath $0^{++}\, 1st\, Radial $&$ t_c$&$\tau$& $\alpha_s$& $m_b$&$G^2$&$G^3$&$G^4$&$M_X$&$f_X$&$ M_{X'}$ [MeV]&$\Delta_{M_{X'}}$&$f_{X'}$ [keV] \\
\hline
\oliva{\it Molecule}\\
${\cal R}_{10}$&149&127&80&93&0&1&0&201&318&20389(441)\\
${\cal R}_{21}$&164&141&73&85&0&1&0&172&179&20403(348)\\
\it Mean &&&&&&&&&&\it \oliva  20398(273) \\
${\cal L}_{0}$& 8&1 &8 &10&0&0.2&0&9&12&--&49&\oliva 138(53)\\
\oliva{\it Four-quark}\\
${\cal R}_{10}$&167&112&88&109&0&2&0&210&300&20491(441)\\
${\cal R}_{21}$&181&127&67&85&0&2&0&179&375&20504(483)\\
\it Mean &&&&&&&&&&\it \purple  20497(326) \\
${\cal L}_{0}$& 8&5 &13 &15&0.6&3&0&14&21&--&77&\purple 203(84)\\
\hline
\end{tabular*}
}
\vspace*{-0.25cm}
    \caption{{\bf Bottom quark channel\,:} The same as in Table\,\ref{tab:psi-psi-0} but for the bottom quark channel.} 
    \label{tab:upsilon-2}
\end{table*}
\subsection*{\b The $2^{++}\,\Upsilon\Upsilon$ molecule 1st radial excitation}
\begin{figure}[hbt]
\begin{center}
\includegraphics[width=7cm]{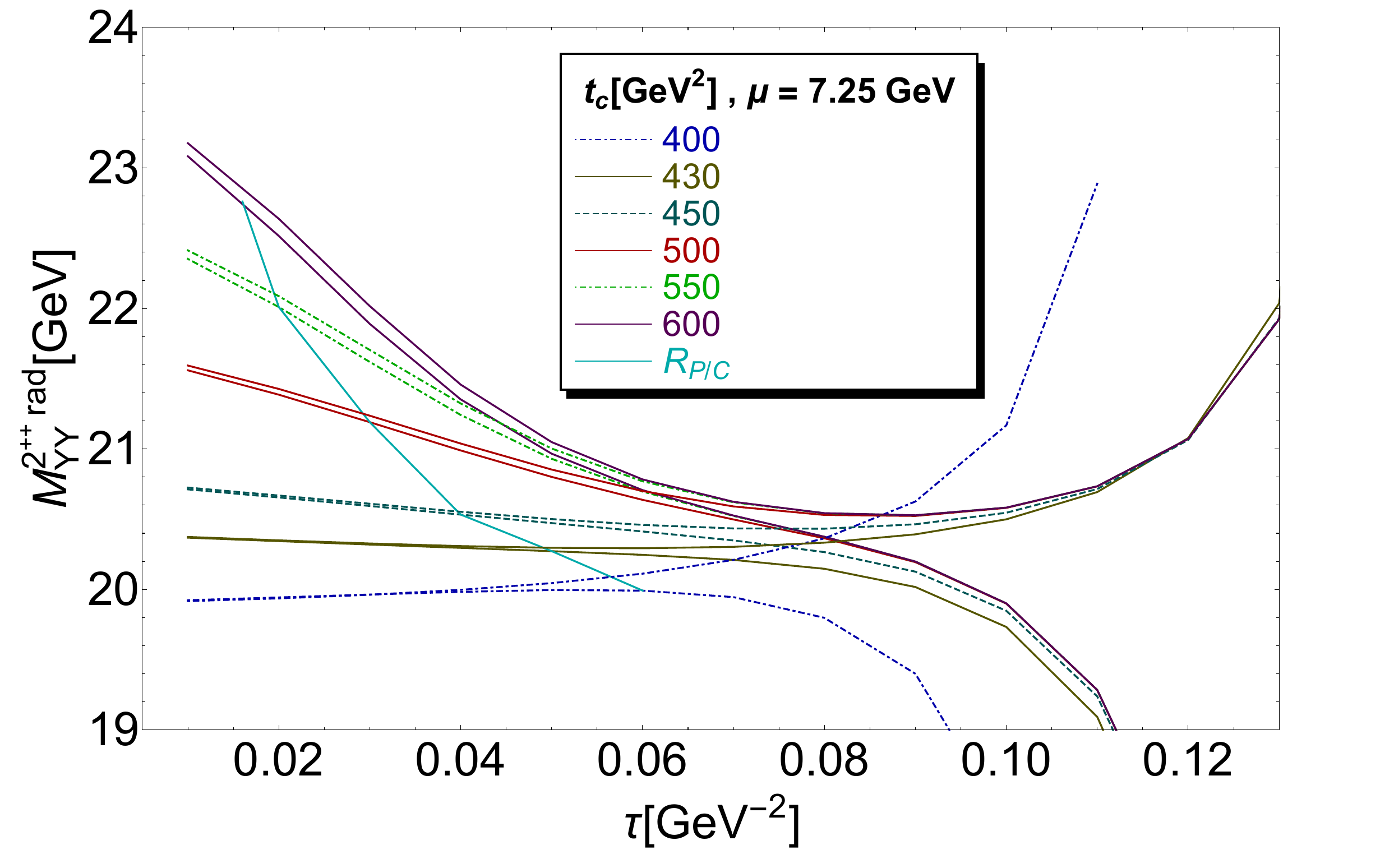}
\vspace*{-0.25cm}
\caption{\footnotesize  The same as in Fig.\,\ref{fig:massb-2} but for the $2^{++} \Upsilon\,\Upsilon$ molecule  radial excitation.}
\label{fig:Upsilon-rad-2}
\end{center}
\vspace*{-0.5cm}
\end{figure}

\d The analysis of the mass from ${\cal R}_{10}$ (curves with inflexion point) and ${\cal R}_{21}$ (curves with minimum)  is shown in Fig.\,\ref{fig:Upsilon-rad-2}. For ${\cal R}_{10}$, the mass values 19992 to 20707 MeV are obtained for $\tau=0.06$ and $t_c$ from 400 to 550 GeV$^2$. For ${\cal R}_{21}$, the mass values 20296 to 20523 MeV are obtained for $(\tau,t_c)$=(0.05,430) to (0.09,500). Their means are quoted in Table\,\ref{tab:upsilon-2} from which we deduce the final prediction. 

\begin{figure}[hbt]
\begin{center}
\includegraphics[width=7cm]{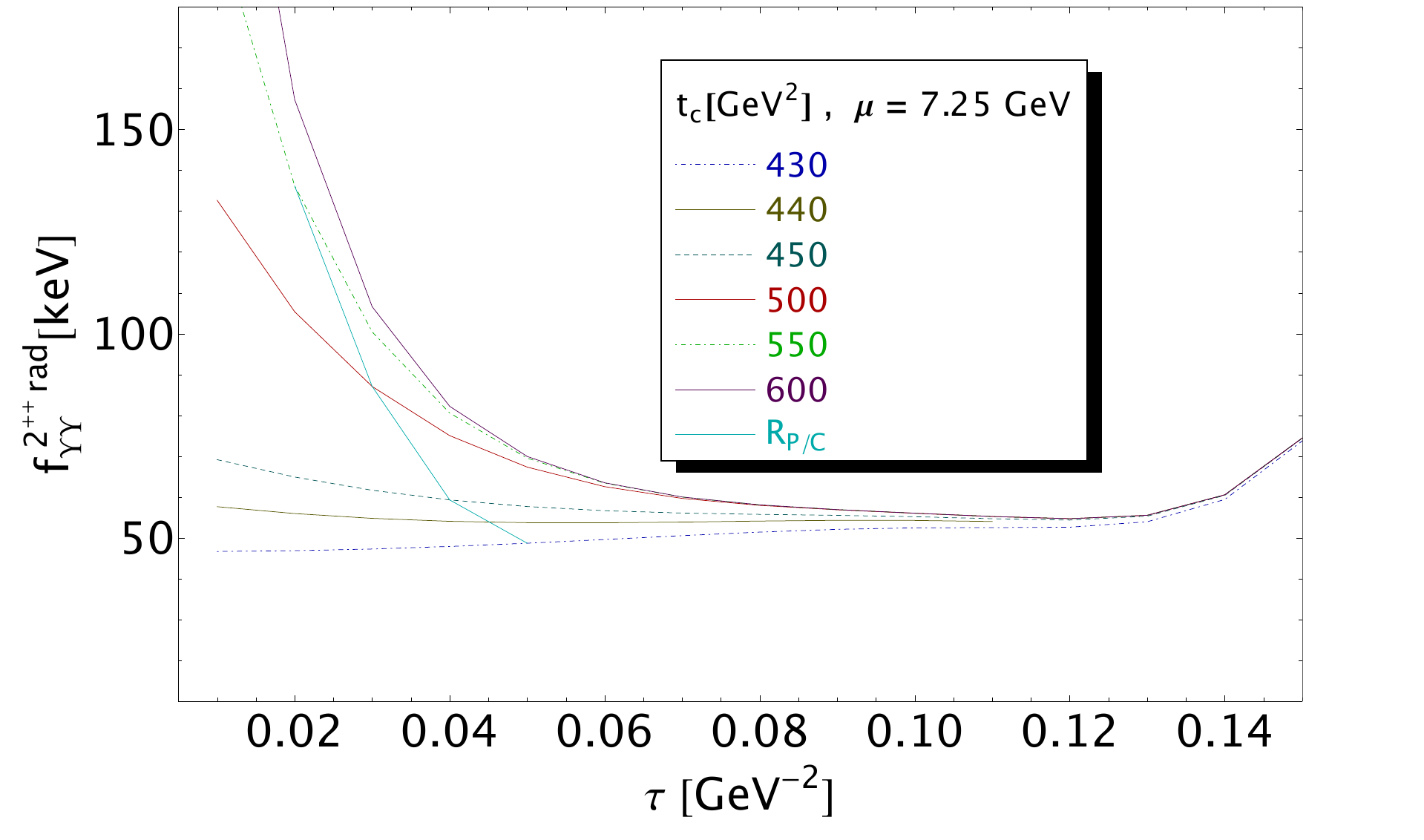}
\vspace*{-0.25cm}
\caption{\footnotesize  The same as in Fig.\,\ref{fig:Upsilon-coupling-2} but for the $2^{++} \Upsilon\,\Upsilon$ molecule  radial excitation.}
\label{fig:Upsilon-coupling-rad-2}
\end{center}
\vspace*{-0.5cm}
\end{figure}

\d The coupling is obtained from ${\cal L}_0$. The curves are shown in Fig.\,\ref{fig:Upsilon-coupling-rad-2}.  $\tau$-stability starts from $(\tau,t_c)$ = (0.06,440) while $t_c$-stability is reached for (0.12,550) to which corresponds $f_{\Upsilon\Upsilon}^{2^{++} {\rm rad}}$ = 53.7 and 54.7 keV. The result is given in Table\,\ref{tab:upsilon-2}. 
\subsection*{\b The $2^{++}\,bb\bar b \bar b$ four-quark  ground state}
The $(\tau,t_c)$ behaviours of the masses and couplings are similar to the one in Fig.\,\ref{fig:massb-2}.

\d For ${\cal R}_{10}$, we have an (almost) $\tau$-stability for $t_c$ = 400\,GeV$^2$ and inflexion point from $(\tau,t_c)$=(0.1,450) corresponding to $M^{2^{++}}_{bb\bar b\bar b}$= 19542 and 19683 MeV. The mean of these two extreme values is quoted in Table\,\ref{tab:upsilon-2}. 

\d ${\cal R}_{21}$ has minimum in $\tau\simeq 0.1$ GeV$^{-2}$ for $t_c\geq 380$ GeV$^2$ and $t_c$-stability for  $(\tau,t_c)$ =(0.22, 450) corresponding to  $M^{2^{++}}_{bb\bar b\bar b}$=19280 and 19386 MeV. The results are given in Table\,\ref{tab:upsilon-2}. 

\d The value of the four-quark state coupling given in Table\,\ref{tab:upsilon-2} is obtained from $(\tau,t_c)$= (0.1,380) to (0.18, 450).
\subsection*{\b The $2^{++}\,bb\bar b \bar b$ four-quark radial excitation}
\d The $(\tau,t_c)$ behaviour of the curves is similar to the ones in Fig.\,\ref{fig:Upsilon-rad-2}. For ${\cal R}_{10}$, the $(\tau,t_c)$ stabilities start from (0.06,400) to (0.06,600) where $M^{2^{++ rad}}_{bb\bar b\bar b}$ = 19992 to 20707 MeV  and for  ${\cal R}_{21}$ from (0.05,430) to (0.09,500) corresponding to $M^{2^{++ rad}}_{bb\bar b\bar b}$ = 20296 and 20523 MeV. Their means are given in Table\,\ref{tab:upsilon-2}.

\begin{figure}[hbt]
\begin{center}
\includegraphics[width=7cm]{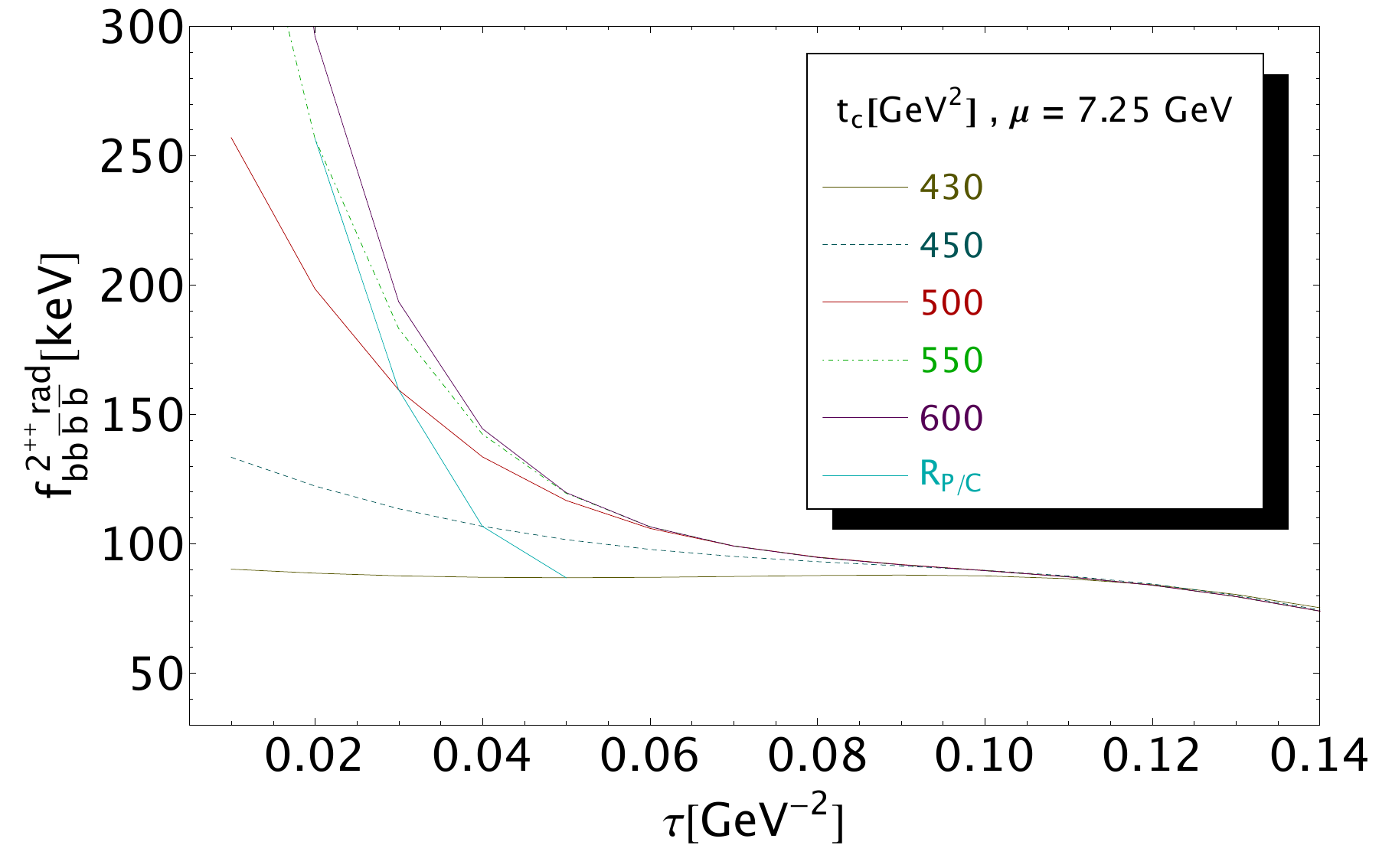}
\vspace*{-0.25cm}
\caption{\footnotesize  The same as in Fig.\,\ref{fig:Upsilon-coupling-rad-2} but for the $2^{++} bb\bar b \bar b$ radial excitation coupling.}
\label{fig:4b-coupling-rad-2}
\end{center}
\vspace*{-0.5cm}
\end{figure}

\d The $(\tau,t_c)$ behaviour of the coupling is shown in Fig.\,\ref{fig:4b-coupling-rad-2}. 
$\tau$ stability starts for $(\tau,t_c)$= (0.05,430) GeV$^2$ to which corresponds $f^{2^{++ rad}}_{bb\bar b\bar b}$=86.8 keV while $t_c$ stability is obtained for (0.09,500) corresponding to $f^{2^{++ rad}}_{bb\bar b\bar b}$=91.9 keV. The result is quoted in Table\,\ref{tab:upsilon-2}.

\subsection*{\b The $0^{++}\,\Upsilon\Upsilon$ molecule and $0^{++}\,bb\bar b \bar b$ ground states}
\begin{figure}[hbt]
\vspace*{-0.25cm}
\begin{center}
\includegraphics[width=7cm]{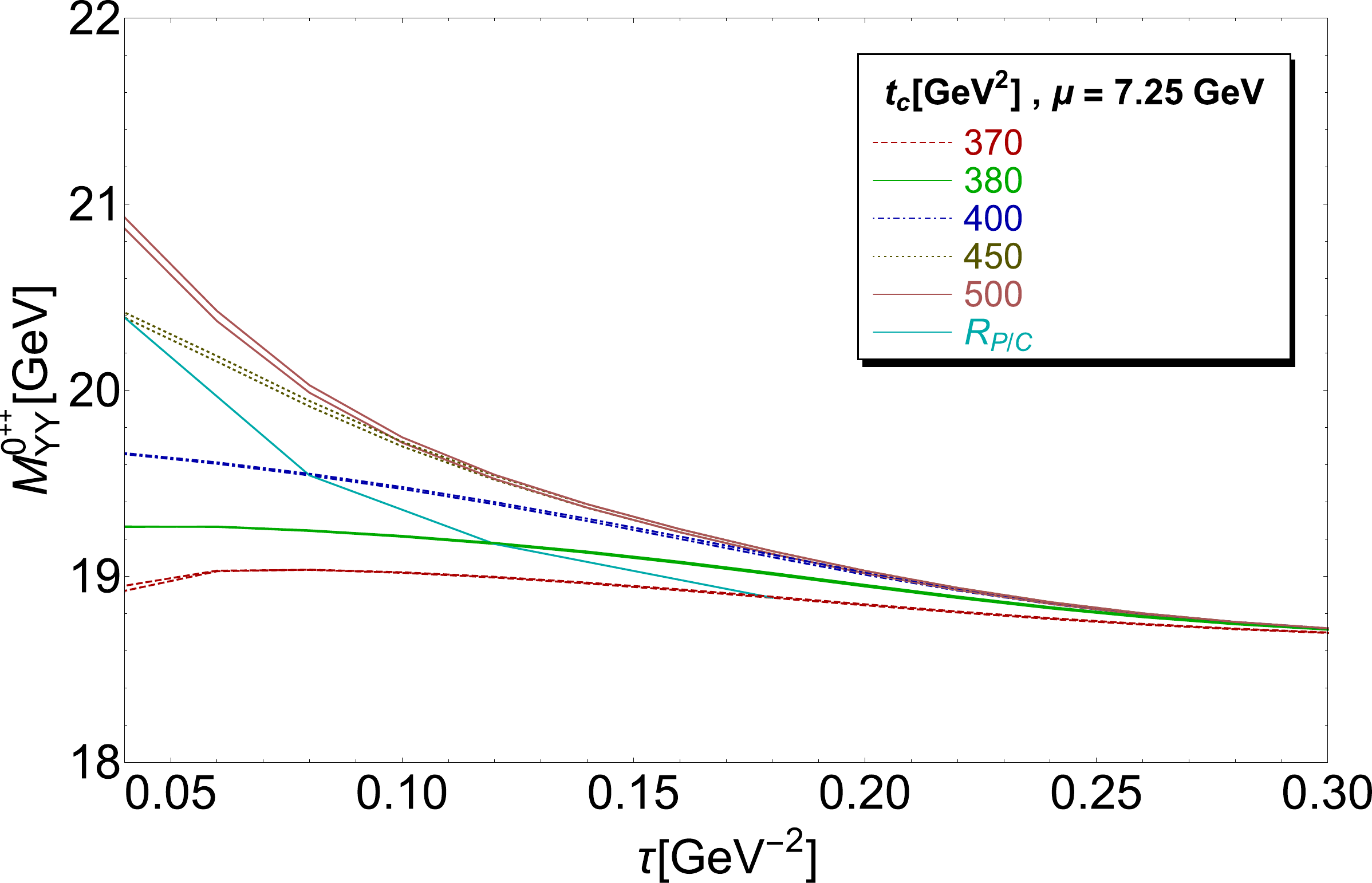}
\vspace*{-0.25cm}
\caption{\footnotesize  The same as in Fig.\,\ref{fig:massb-2} but for the $0^{++}\,\Upsilon\Upsilon$ molecule  mass.}
\label{fig:Upsilon-mass-0}
\end{center}
\vspace*{-0.5cm}
\end{figure} 

\d  The analysis of the mass is similar to the case of the $2^{++}$ states. We show, in Fig.\,\ref{fig:Upsilon-mass-0}, the different curves in the case of the $0^{++}\,\Upsilon\Upsilon$ molecule for ${\cal R}_{10}$ and ${\cal R}_{21}$ which overlap each others. The $\tau$ stability for  $\tau\leq 0.13$ GeV$^{-2} $ and $t_c=370$ GeV$^2$ is excluded by the  $R_{P/C}$ condition. Our optimal results are obtained for $\tau\simeq 0.13$ GeV$^{-2} $ and $t_c=380$ to 450 GeV$^2$ to which corresponds $M^{0^{++}}_{\Upsilon\Upsilon}$ = 19280 to 19414 MeV for ${\cal R}_{10}$ and 19156 to 19463 for   ${\cal R}_{21}$. Final results are quoted in Table\,\ref{tab:upsilon-2}.

\d Similar results are obtained for the $bb\bar b\bar b$ four-quark states. Optimal results are obtained for $\tau\simeq 0.12$ GeV$^{-2} $ and$t_c$ from 380 to 450 GeV$^2$ to which corresponds $M^{0^{++}}_{bb\bar b \bar b}$ = 19170 to 19526 MeV for  ${\cal R}_{10}$ and 19177 -- 19550 MeV for ${\cal R}_{21}$.  Final results are quoted in Table\,\ref{tab:upsilon-2}.

\begin{figure}[hbt]
\vspace*{-0.25cm}
\begin{center}
\includegraphics[width=7cm]{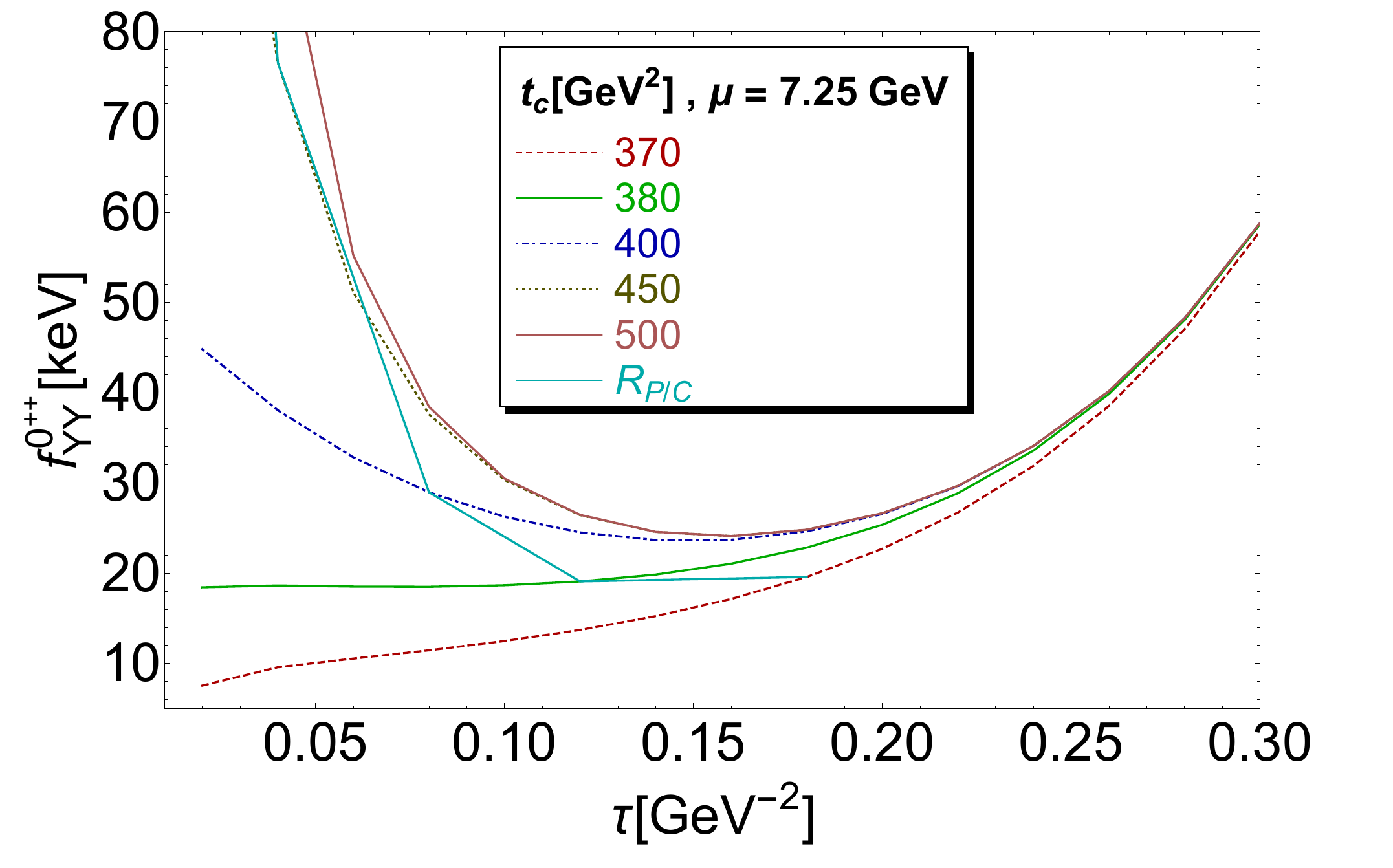}
\vspace*{-0.25cm}
\caption{\footnotesize The same as in Fig.\,\ref{fig:psi-psi-coupling} but for  the $0^{++} \Upsilon\,\Upsilon$ molecule  coupling.}
\label{fig:Upsilon-f-0}
\end{center}
\vspace*{-0.5cm}
\end{figure}

\d The analysis of the coupling  $f^{0^{++}}_{\Upsilon\Upsilon}$  is shown in Fig.\,\ref{fig:Upsilon-f-0}.  Stability in $\tau$ and $t_c$ is reached for $(\tau,t_c)$ = (0.12,390) and (0.14,450) which corresponds $f^{0^{++}}_{\Upsilon\Upsilon}$= 25.05 and 27.83 keV. The final result is given in Table\,\ref{tab:upsilon-2}.

\d The curves for $f^{0^{++}}_{bb\bar b \bar b}$ are similar to the ones of $f^{0^{++}}_{\Upsilon\Upsilon}$. Stabilities are reached for  $(\tau,t_c)$ =(0.10,390) and (0.14,450) corresponding to $f^{0^{++}}_{bb\bar b \bar b}$ = 38.72 and 43.12 keV. 
The final result is given in Table\,\ref{tab:upsilon-2}.

\subsection*{\b The $0^{++}\,\Upsilon\Upsilon$ molecule and $bb\bar b \bar b$ 1st radial excitations}
For the masses, the curves are similar to the case of the $2^{++}$ shown in Fig.\,\ref{fig:Upsilon-rad-2}. 

\d In the case of molecule, ${\cal R}_{10}$, stabilizes for (0.07,430) and (0.07,550) to which corresponds to $M_{\Upsilon\Upsilon}^{0^{++} {\rm rad}}$= 20239 and 20538 MeV. For ${\cal R}_{21}$, stabilities are obtained for (0.08,430) and (0.07,550) which corresponds to $M_{\Upsilon\Upsilon}^{0^{++} {\rm rad}}$ = 20239 and 20567 MeV.
Final results are given in Table\,\ref{tab:upsilon-2}.

\d  In the case of $bb\bar b \bar b$,  ${\cal R}_{10}$, stabilizes for (0.05,430) and (0.07,550) which corresponds to $M^{0^{++}}_{bb\bar b \bar b}$= 20324 and 20658 MeV. ${\cal R}_{21}$, stabilizes for (0.06,430) and (0.07,550) corresponding to $M^{0^{++}}_{bb\bar b \bar b}$=20323 and 20684 MeV. Final results are quoted in Table\,\ref{tab:upsilon-2}.

\d The couplings are obtained from ${\cal L}_0$. They behave as in Fig.\,\ref{fig:Upsilon-f-0}. For the molecule, stability is reached for (0.04,440) and (0.09,550) corresponding to $f_{\Upsilon\Upsilon}^{0^{++} {\rm rad}}$ = 129.8 and 145.4 keV. For the four-quark state, one has stability for (0.04,440) and (0.12,550) corresponding to 
$f^{0^{++ rad}}_{bb\bar b\bar b}$ = 194.8 and 210.2 keV.  Final results are quoted in Table\,\ref{tab:upsilon-2}.

\section{ Comments and comparison with some QSSR results}

Besides some eventual mistakes in the calculation of the spectral function by different groups which are not easy to check due to the complexity of the QCD calculation, we give some general remarks on the papers in the literature. 
\subsection*{\b Charm quark channel}
\begin{table}[hbt]
\vspace*{-0.5cm}
\setlength{\tabcolsep}{0.85pc}
\newlength{\digitwidth} \settowidth{\digitwidth}{\rm 0}
    {\scriptsize\begin{tabular}{lllll}

&\\
\hline
$M_X$ [MeV]&$f_X$ [keV]& $t_c$ [GeV$^2$] & $\tau$ [GeV$^{-2}$]&Refs.    \\
\hline
\boldmath$2^{++}$\\
{\it\oliva Molecule} \\
\rowcolor{yellow} 6521(69) &96(14)&$45,\, 70$& Voir text&This work \\
6250 &&$45\sim 46$ &$0.19\sim 0.21$&\cite{AZIZI1}\\

{\it \purple Four-quark} \\
\rowcolor{yellow}6347(99) &151(29)&$45,\, 80$& Voir text&This work \\
$6090(80)$ && $44(1)$ & $0.19\sim 0.22$&\cite{WANG}\\
$6609(50)$ &&$51- 52$ &$0.15\sim 0.18$&\cite{AZIZI2}\\
$6370(190)$ &&-- &Moments&\cite{ZHU}\\

{\it \blue Mixed States} \\
$6320(110)$ &&$50(1)$ &$0.41(0.01)$&\cite{RAPHAEL}\\
\hline
\boldmath $0^{++}$\\
{\it\oliva Molecule} \\
\rowcolor{yellow} 6394(115)&194(39)&$45,\, 80$&Voir text& This work \\
6250 &&$45\sim 46$ &$0.19\sim 0.22$&\cite{AZIZI1}\\

{\it \purple Four-quark} \\
\rowcolor{yellow} 6380(112)&303(57)&$45,\, 80$&Voir text& This work \\
$5990(80)$ && $42(1)$ & $0.22\sim 0.24$&\cite{WANG}\\
$6590(170)$ &&-- &Moments&\cite{ZHU}\\
\hline
\end{tabular}
}
\vspace*{-0.25cm}
    \caption{$2^{++}$ and $0^{++}$ $J/\psi J/\psi$ molecule and $cc\bar c\bar c$ four-quark ground states masses predicted by different groups using Laplace  and Moment sum rules.  We refrain to average these results as most of them do not satisfy the $R_{P/C}$ requirement given in Eq.\,\ref{eq:rpc}.} 
    \label{tab:mass}
\end{table}

\d We notice that the expressions of the spectral functions used in the literature are often  to LO of PT. However, we have (repeatedly) mentioned in our different works\,\cite{HEP18,SU3,QCD16,MOLE16,X5568,MOLE12,MOLE20} that the definition of the heavy quark mass is ill-defined , at LO of PT. Many authors use as input the value of the running $\overline{MS}$ heavy quark mass which is unjustified as the spectral function is computed at LO using (without saying) the on-shell scheme where the on-shell or pole quark mass appears implicitly. None of the works in the QSSR literature discuss the systematic errors induced by a such ill-defined mass definition at LO as emphasized in our previous works. To circumvent this problem, we have included the factorised NLO PT contributions using the convolution of two bilinear $\bar cc$ two-point functions\,\cite{PICH,SNPIVO} and replaced the pole mass appearing in the expression of the spectral function by the running $\overline{MS}$ through their known relation to NLO\,\cite{SNB1,SNB2}. Though the NLO contribution is not large, it gives some meaning of the heavy quark input mass values used. 

\d All authors working with LSR use the ratio of moments ${\cal R}_{10}$ shown in  Fig.\,\ref{fig:psi-psi-mass} and the MDA parametrization of the spectral function for determining the lowest ground state masses.  We notice that some of them use a relatively low  and narrow range values of the QCD continuum threshold $t_c\equiv s_0\simeq (45-50)$ GeV$^2$ at the beginning of the $\tau$ stability ($\tau\equiv 1/M_0^2$  below  0.24 GeV$^{-2}$) (see Table\,\ref{tab:mass}).
However, most of these low sets of $(\tau,t_c)$ is excluded by the $R_{P/C}$ requirement given in Eq.\,\ref{eq:rpc}.  It is also clear from Fig.\,\ref{fig:psi-psi-mass}, that in this range of $\tau$-values the role of the QCD continuum contribution is important and a slight variation of the $t_c$ affects strongly the prediction of the resonance mass such that the mass value is often underestimated and unreliable. 

\d To check and to supplement the results obtained from 
${\cal R}_{10}$, 
we also use the higher ratio ${\cal R}_{21}$. Its exhibits $\tau$-minimum  and $t_c$-stability and gives almost the same results as ${\cal R}_{10}$ (see Fig.\,\ref{fig:psi-psi-mass}). 


\d Our results for the ground state masses quoted in Table\,\ref{tab:mass} obtained for a large range of $t_c$-values and which satisfies the $R_{P/C}$ requirement are more conservative and is less dependent on the choice of $t_c$-value contrary to the estimates of some authors
where the $t_c$ values have been often ajusted to be around an empirical value of the first radial excitation masses extrapolated from the known meson ones . However, as the QCD continuum contribution smears all contribution of higher states contributions to the spectral integral, it is not (a priori) clear that the value of $t_c$ co\"\i ncide with the mass of the first radial excitation.


\d In the case of $2^{++}$ channel and using the same set of $(\tau,t_c)$ parameters, we do not recover the result of Ref.\,\cite{AZIZI1} for the molecule state. For the four-quark state, we do not recover the result of Ref.\,\cite{WANG} which is too low while the one of \cite{AZIZI2} is too high.  Our result for the four-quark state is in perfect agreement  with the one from moment sum rules where the higher states and QCD continuum contributions are neglected\,\cite{ZHU}.  Comparing our result with the one in Ref.\,\cite{RAPHAEL} is quite delicate as the author predicts an averaged molecule and four-quark masses found to mix via an angle $\theta\approx 0.17^0$ in the QCD side.  However, we fail to understand how the meson couplings which are not equal disappear in the LHS of ${\cal R}_{10}$ used to get the masses.

\d In the case of the $0^{++}$ channel and using the same set of $(\tau,t_c)$ parameters, we recover within the errors the LO results of Ref.\,\cite{AZIZI1} for the molecule state.
For the four-quark state, the result of \,\cite{WANG} looks to be too low, while we agree within the errors with the one from moment sum rules where the higher states and QCD continuum contributions are neglected\,\cite{ZHU}. 

\d We compare the results for the $0^{++}$ with the previous ones in Ref.\,\cite{MOLE20}. We notice that though the two results agree within the errors, the central values differ by 178 MeV where in Ref.\,\cite{MOLE20} we have only considered the ones at the inflexion point which also explains the small error due to $t_c$ in Ref.\,\cite{MOLE20}. We also mention that the quoted value in Ref.\,\cite{MOLE20} is the mean between the one where the OPE is truncated  up to the $\la \alpha_s G^2\ra$ with the one up to $\la g^3_s G^3\ra$  condensates contributions. 

\d  Our results in Table\,\ref{tab:psi-psi-2} indicate that the mass-splittings of the $2^{++}$ and $0^{++}$ states between the 1st radial excitations and the ground states are in the range (1392-1466) MeV which is about 2.5 times $M_{\psi (2S)} - M_{J/\psi J/\psi}\simeq 650$ MeV.   We also note that the couplings of the 1st radial excitations to the  interpolating currents are about 3 times the ones of the ground states,  which is unusual.
These peculiar features may indicate some new dynamics of the molecule and four-quark states.
\subsection*{\b Bottom quark channel}
\begin{table}[hbt]
\vspace*{-0.5cm}
\setlength{\tabcolsep}{0.55pc}
    {\scriptsize
\begin{tabular}{llllll}

&\\
\hline
$M_X$ [MeV]&$\tau$ [GeV$^{-2}$]&$f_X$ [keV]& $\tau$ [GeV$^{-2}$]& $t_c$ [GeV$^2$] &Refs.    \\
\hline
\boldmath$2^{++}$\\
{\it\oliva Molecule} \\
\rowcolor{yellow} 19332(72) &See text&12.3(3.4)&See text&$380,\, 450$&This work \\

{\it \purple Four-quark} \\
\rowcolor{yellow}19306(54) &See text&16.1&See text&$380,\, 450$&This work \\
$18850(90)$ && & $0.07$&$375(3)$ &\cite{WANG}\\
$18530(86)$ &&&0.06&$380-385$ &\cite{AZIZI3}\\
$18320(170)$ &&-- &Moments&--&\cite{ZHU}\\
\hline
\boldmath $0^{++}$\\
{\it\oliva Molecule} \\
\rowcolor{yellow} 19302(123)&0.13&21.6& &$380,\,450$& This work \\

{\it \purple Four-quark} \\
\rowcolor{yellow} 19355(146)&0.12&$29.1$& &$380,\, 450$& This work \\
$18840(90)$ &&& $0.07$& 374(3) &\cite{WANG}\\
$18590(170)$ &&-- &Moments&--&\cite{ZHU}\\
\hline
\end{tabular}
}
\vspace*{-0.25cm}
    \caption{$2^{++}$ and $0^{++}$ $\Upsilon\Upsilon$ molecule and $bb\bar b\bar b$ four-quark states masses predicted by different groups using Laplace  and Moment sum rules.} 
    \label{tab:massb}
\end{table}

\d In the case of the $2^{++}$ four-quark state, the result of Ref.\,\cite{WANG} taken at the same value of $t_c$ agrees with ours. However,  the mass increases with $t_c$ until the stability region reached for $t_c\geq 450$ GeV$^2$. The results of Refs.\,\cite{AZIZI3,ZHU} are too low and outside the $\tau$-stability region. 

\d In the case of the $0^{++}$ channel, we notice that the errors given by Refs.\,\cite{WANG,ZHU} are relatively small compared to ours which come mainly from the large range of $t_c$ used in this work, while the central values given by Refs.\,\cite{WANG,ZHU} are much lower than ours. 
The comments on the comparison in the charm quark channel with Ref.\,\cite{MOLE20} also apply in this bottom quark channel. 
\section{Some phenomenological implications}
\subsection*{\b Charm quark channel}

\d The predicted $2^{++}$ molecule lowest ground state mass of  6521(69) MeV may favour the molecule interpretation of the $X(6600)$, while the four-quark mass prediction  is about 3.7 $\sigma$ below the experimental candidate.

\d Using a two-component mixing scheme, one may also interpret  the  X(6900) and X(7100) as a mixture of the four-quark ground state with its1st radial excitation  via an  angle $\theta\approx 5^0$.

\d  The masses of the $2^{++}$ and $0^{++}$ ground states are (almost) degenerated (one has the same feature for the 1st radial excitations) such that it may be difficult to disentangle these states from the data without some careful partial waves analysis. 

\subsection*{\b Bottom quark channel}
Our results in Table\,\ref{tab:upsilon-2} indicate that the lowest ground states masses of the molecule and four-quark $2^{++}$ and $0^{++}$ states are (almost) degenerated. The same feature is observed for their 1st radial excitations indicating that separating these states from the data is non-trivial and again need a careful partial-wave analysis if the mass region is accessible.


\end{document}